\documentclass[times]{aastex631}
\usepackage{amssymb}
\usepackage{amsmath}
\usepackage{gensymb}
\usepackage{graphicx}
\usepackage{xcolor}
\usepackage{natbib}
\usepackage{multirow}
\newcommand{\comment}[1]{}

\newcommand{\beq}{\begin{equation}}
\newcommand{\eeq}{\end{equation}}

\newcommand{\Ms}{\textrm{M}_*}
\newcommand{\Msun}{\textrm{M}_\odot}
\newcommand{\kmps}{km~s$^{-1}$}
\newcommand{\MHI}{\rm{M_{H\textsc{i}}}}
\newcommand{\MHtwo}{\rm{M}_{\rm H_2}}

\newcommand{\tdephi}{{\rm t_{dep,H{\textsc i}}}}
\newcommand{\tdephtwo}{{\rm t_{\rm dep,H_2}}}

\newcommand{\htwo}{{\rm H_2}}

\newcommand{\hi}{H{\sc i}}

\newcommand{\hii}{H{\sc i} 21\,cm}
\defcitealias{NatPaper}{vD18}
\graphicspath{{./}{figures/}}

\shorttitle{The GMRT-CAT$z1$ Survey}
\shortauthors{Chowdhury, Kanekar and Chengalur}

\begin{document}
	
	\title{The Giant Metrewave Radio Telescope Cold-HI AT $z\approx1$ Survey}

	\correspondingauthor{Aditya Chowdhury}
	\email{chowdhury@ncra.tifr.res.in}
	
	\author{Aditya Chowdhury}
	\affil{National Centre for Radio Astrophysics, Tata Institute of Fundamental Research, Pune, India.}
	
	\author{Nissim Kanekar}
	\affil{National Centre for Radio Astrophysics, Tata Institute of Fundamental Research, Pune, India.}
	
	\author{Jayaram N. Chengalur}
	\affil{National Centre for Radio Astrophysics, Tata Institute of Fundamental Research, Pune, India.}

	
	
	\begin{abstract}
  We describe the design, data analysis, and basic results of the Giant Metrewave Radio Telescope Cold-H{\sc i} AT $z\approx1$ (GMRT-CAT$z1$) survey,  a $510$-hour upgraded GMRT H{\sc i} 21\,cm emission survey of galaxies at $z=0.74-1.45$ in the DEEP2 survey fields. The GMRT-CAT$z1$ survey is aimed at characterising  H{\sc i}\ in galaxies during and just after the epoch of peak star-formation activity in the Universe, a key epoch in galaxy evolution. We obtained high-quality H{\sc i} 21\,cm spectra for 11,419 blue star-forming galaxies at $z=0.74-1.45$, in seven pointings on the DEEP2 subfields. We detect the stacked H{\sc i} 21\,cm emission signal of the 11,419 star-forming galaxies, which have an average stellar mass of $\Ms\approx10^{10}~\Msun$,  at $7.1\sigma$ statistical significance, obtaining an average H{\sc i}\ mass of $\langle\rm{M_{H{\textsc i}}}\rangle=(13.7\pm1.9)\times10^{9}~\textrm{M}_\odot$. This is significantly higher than the average H{\sc i} mass of $\langle\rm{M_{H{\textsc i}}}\rangle=(3.96 \pm 0.17)\times10^{9}~\textrm{M}_\odot$ in star-forming galaxies at $z \approx 0$ with an identical stellar-mass distribution. We stack the rest-frame 1.4~GHz continuum emission of our 11,419 galaxies to infer an average star-formation rate (SFR) of $8.07\pm0.82~\textrm{M}_\odot\textrm{yr}^{-1}$. Combining our average H{\sc i} mass and average SFR estimates yields an H{\sc i}\ depletion timescale of  $1.70\pm0.29$~Gyr, for star-forming galaxies at $z\approx1$, $\approx3$ times lower than that of local galaxies. We thus find that, although main-sequence galaxies at $z\approx1$ have a high H{\sc i} mass, their short H{\sc i}\ depletion timescale is likely to cause quenching of their star-formation activity in the absence of rapid gas accretion from the circumgalactic medium.
	\end{abstract}
	
	
   \section{Introduction}
Over the past seven decades, the \hii\ transition has played a fundamental role in our understanding of galaxies and galaxy evolution. Studies of  \hii\ emission from galaxies in the local Universe, using both single dishes \citep[e.g.][]{Zwaan05,Haynes18,Catinella18} and radio interferometers \citep[e.g.][]{Verheijen01,Begum08,Walter08,Sera11,Hunter12}, have provided information on the nature of dark matter halos in nearby galaxies \citep[e.g.][]{Bosma81,Casertano91,deBlok08,Begum08}, the mass distribution of \hi\ in the local Universe \citep[the \hi\ mass function; e.g.][]{Zwaan97,Zwaan05,Jones18}, the connection between rotation velocity and baryonic mass \citep[e.g.][]{Tully77,McGaugh00,Verheijen01,Lelli19}, the dependence of the global \hi\ properties of galaxies on their stellar properties \citep[e.g.][]{Saintonge22}, the relation between the \hi\ mass and the size of the \hi\ disk \citep[e.g.][]{Broeils97,Wang16}, the role of \hi\ in regulating star-formation in galaxies \citep[e.g.][]{Leroy08,Roychowdhury09}, the impact of environment and galaxy interactions on the \hi\ disks of galaxies \citep[e.g.][]{Cayatte90,Hibbard01,Chung09}, the cosmological mass density of neutral hydrogen \citep[e.g.][]{Zwaan05,Jones18}, 
and a variety of other issues.

Unfortunately, the low Einstein A-coefficient of the \hii\ transition implies that the \hii\ line is very weak. This makes \hii\ emission studies of galaxies at cosmological distances challenging, even with sensitive modern-day telescopes. Indeed, while deep searches for \hii\ emission at high redshifts have been carried out for more than thirty years
\citep[e.g.][]{Subrahmanyan90,Uson91,Wieringa92}, often with integrations of hundreds of hours  \citep[e.g.][]{Jaffe13,Catinella15,Fernandez16,Gogate20}, the highest redshift at which \hii\ emission has been detected from an individual object remains $z \approx 0.376$ \citep{Fernandez16}. Detecting \hii\ emission from galaxies at significantly higher redshifts, $z \gtrsim 1$, is one of the main goals of next-generation telescopes like the Square Kilometre Array.

The prohibitively-large integration times required on today's radio telescopes to detect \hii\ emission from  galaxies at $z\gtrsim1$ has meant that, until very recently, nothing was known about the \hi\ masses of high-$z$ galaxies or their evolution, or relations to other galaxy properties such as stellar masses, SFR, molecular gas mass, environment, etc. This lack of information about \hi, the primary fuel for star-formation, in high-$z$ galaxies is in stark contrast with the spectacular progress made in measuring their stellar properties \citep[e.g.][]{Madau14}, and, more recently, their molecular gas properties \citep[e.g.][]{Tacconi20}.

Progress in understanding the \hi\ properties of high-$z$ galaxies can be made by using the \hii\ stacking approach \citep{Zwaan00,Chengalur01}. Here, the \emph{average} \hi\ properties of a sample of galaxies can be determined by co-adding, i.e. ``stacking'', the \hii\ emission signals of the individual galaxies, as long as their positions and redshifts are accurately known. Applying this technique to determine the average \hi\ mass of galaxies at $z \gtrsim 1$ requires a large number of galaxies with spectroscopic redshifts, located within the field of view of the telescope, and with their redshifted \hii\ line frequencies covered by a single frequency setting. Further, \hii\ stacking experiments critically require accurate spectroscopic redshifts, with redshift errors $\lesssim 100$~\kmps , to prevent the stacked signal from being smeared in velocity \citep[e.g.][]{Maddox13,Elson19}.

In the local Universe, \hii\ stacking has been used to probe the dependence of the \hi\ properties of galaxies on their stellar properties and environments \citep{Fabello11,Fabello12,Brown15,Brown17,Meyer16}, as well as to measure the cosmological \hi\ mass density at $z\approx0$ \citep[e.g.][]{Hu20}.  The results of these studies have been consistent with those from direct \hii\ emission surveys of individual galaxies. At low redshifts, $z\lesssim 0.2$, successful \hii\ stacking experiments have been carried out using both single dishes \citep{Delhaize13} and radio interferometers \citep{Rhee13}. At intermediate redshifts, $z\approx0.2-0.4$, the early \hii\ stacking experiments, all using the Giant Metrewave Radio Telescope (GMRT),  obtained only tentative ($\approx2-3~\sigma$) detections of the stacked \hii\ emission signal \citep{Lah07,Lah09,Rhee16,Rhee18}. Recently, \citet{Bera19} used a 175-hr upgraded GMRT\footnote{The GMRT was upgraded in 2018 with new wideband recievers and a wideband correlator. The upgrade increased the maximum instantaneous bandwidth from 33.3~MHz to 400~MHz.} survey of the Extended Groth Strip at $z \approx 0.2-0.4$ to obtain the first statistically-significant detection of the stacked \hii\ emission from galaxies at intermediate redshifts.  At even higher redshifts, \citet{Kanekar16} used 60~hrs with the GMRT to search for the stacked \hii\ emission from star-forming galaxies at $z\approx1.3$ in the DEEP2 survey fields \citep{Newman13}, obtaining an upper limit on the average \hi\ mass of the galaxies.

Recently, \citet{Chowdhury20} used the upgraded GMRT to carry out an \hii\ emission survey of galaxies in the DEEP2 survey fields \citep{Newman13}, covering the redshift range $z = 0.74-1.45$. Chowdhury et al. stacked the \hii\ signal from 7,653 blue star-forming galaxies at $z = 0.74-1.45$ to measure, for the first time,  the average \hi\ mass of galaxies at $z\approx1$. They found that the average \hi\ mass of blue star-forming galaxies at $z \approx 1$, with an average stellar mass of $\Ms\approx10^{10}~\Msun$, is a factor of $\approx3$ higher than that of galaxies at $z \approx 0$ with the same average stellar mass. However, they also found that the \hi\ reservoir can sustain the high SFR in these galaxies for only $\approx1-2$~Gyr, in the absence of fresh gas accretion. Subsequently, \citet{Chowdhury21} reported results from another GMRT \hii\ emission survey of the DEEP2 fields, using the original GMRT receivers and a 33-MHz bandwidth to cover the narrow redshift range $z=1.18-1.39$. They stacked the \hii\ emission from 2,841 blue star-forming galaxies to obtain an independent $\approx 5\sigma$ detection of the stacked \hii\ signal, at $z \approx 1.3$. Again, the \hi\ reservoir was found to be sufficient to sustain the galaxy SFRs for only $\approx 2$~Gyr, in the absence of fresh gas accretion. \citet{Chowdhury20} suggested that the observed decline in the star-formation activity of the Universe at $z\lesssim1$ arises due to insufficient gas accretion on to star-forming galaxies from the circumgalactic medium \citep[CGM; see also][]{Bera18,Chowdhury21}.

We present here the GMRT Cold-\hi\ AT $z\approx1$ (GMRT-CAT$z1$) survey, a 510-hr upgraded GMRT  \hii\ emission survey of galaxies at $z=0.74-1.45$ in the DEEP2 survey fields, aimed at characterising the \hi\ properties of galaxies at $z \approx 1$. The survey covers a crucial epoch where the cosmic SFR density of the Universe begins to decline after its peak at $z \approx 1-3$ \citep[e.g.][]{Madau14}; the GMRT-CAT$z1$ survey is thus uniquely suited to investigate the role of \hi\ in the decline of the SFR density at $z\lesssim1$.  This paper describes the survey design, the observations, and the data analysis of the GMRT-CAT$z1$ survey, along with an accurate measurement of the average \hi\ mass and the \hi\ depletion timescale of star-forming galaxies at $z\approx1$.  A study of the role of \hi\ in the decline of the cosmic SFR density is presented in \citet{Chowdhury22}, while \citet{Chowdhury22b} compare the contributions of \hi, $\htwo$, and stars to the baryonic content of star-forming galaxies at $z\gtrsim 1$. Future papers will discuss other results from the GMRT-CAT$z1$ survey, including the \hi\ scaling relations at these redshifts, a comparison between the properties of the ionized gas and the neutral atomic gas in star-forming galaxies at $z\approx1$, and the cosmological \hi\ mass density at $z\approx1$.

 This paper is organised as follows: Section~\ref{sec:deep2} provides a brief summary of the DEEP2 Galaxy Redshift Survey, focussing on the aspects relevant for our GMRT \hii\ survey; Section~\ref{sec:obs} discusses the design of the GMRT-CAT$z1$ survey and provides information on the upgraded GMRT observations; Section~\ref{sec:analysis} presents the analysis of the GMRT data in detail;
Section~\ref{sec:stacking} describes the procedures used to stack the \hii\ emission luminosities and the rest-frame 1.4~GHz continuum luminosities; Section~\ref{sec:sample} provides detailed information on the main sample of galaxies; Section~\ref{sec:massresolution} presents our measurements of the average \hi\ mass as a function of spatial resolution and our final choice of the optimal spatial resolution for all subsequent \hii\ stacking;  Section~\ref{sec:histack} presents the GMRT-CAT$z1$ detection of the stacked \hii\ line emission and stacked rest-frame 1.4~GHz continuum emission from our full sample of galaxies at the optimal spatial resolution, along with the main results of this paper concerning the average \hi\ properties of star-forming galaxies at $z \approx 1$; { Section~\ref{sec:redagn} provides a discussion on the average \hi\ properties of red galaxies and galaxies hosting active galactic nuclei (AGNs) at $z\approx1$}; Section~\ref{sec:systematic} discusses possible systematic effects, and shows that our average \hi\ mass measurements are unlikely to be affected by such systematics; Section~\ref{sec:uplims} presents a search for \hii\ emission from the individual DEEP2 galaxies covered by the GMRT-CAT$z1$ survey;  
and, finally,  Section~\ref{sec:conclusion} summarises our key conclusions. We refer the reader who is interested in the key science results of this paper to Sections~\ref{sec:histack} and \ref{sec:conclusion}.

Throughout this paper, we use a flat ``737'' $\Lambda$-cold dark matter ($\Lambda$CDM) cosmology, 
with ($\rm H_0$, $\rm \Omega_{m}$, $\rm \Omega_{\Lambda})=(70$~km~s$^{-1}$~Mpc$^{-1}$, $0.3, 0.7)$. We also assume a Chabrier initial mass function (IMF) in all estimates of stellar masses and SFRs. SFR estimates from the literature that assume a Salpeter IMF were converted to a Chabrier IMF by subtracting 0.2~dex \citep[e.g.][]{Madau14}. All magnitudes are in the AB system. 

\section{The DEEP2 Galaxy Redshift Survey}
\label{sec:deep2}
\begin{figure}
    \centering
    \includegraphics[width=\linewidth]{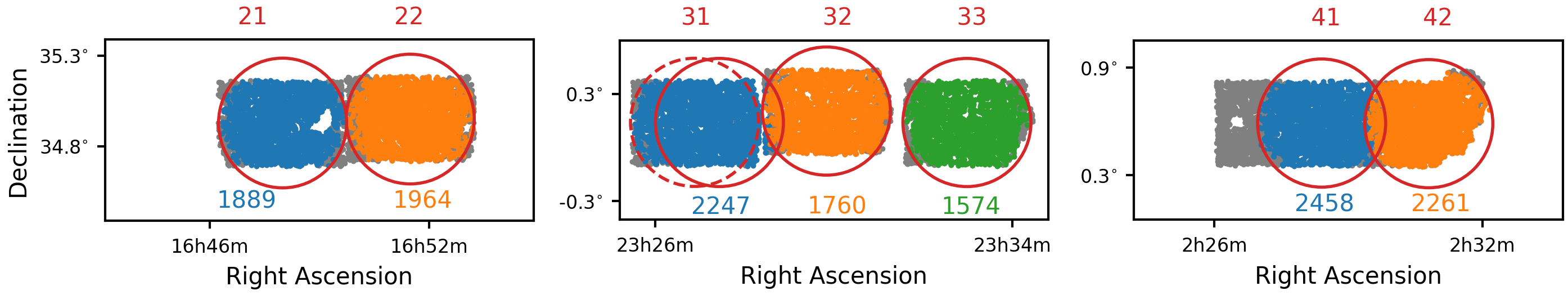}
    \caption{The sky area covered by the GMRT-CAT$z1$ survey in the DEEP2 survey fields 2, 3, and 4. The three panels show the three DEEP2 fields, with the regions covered by the DEEP2 survey marked in gray. The red circles show the half-power point of the GMRT primary beam at $\approx 610$~MHz. The regions in blue, orange, and green show the galaxies covered by the GMRT-CAT$z1$ survey, with the number of blue galaxies in each pointing shown below each circle, and the sub-field number indicated above each circle. The dashed circle shows a slightly different pointing on DEEP2 sub-field 31 that was used in the observations of Cycle~35.}
    \label{fig:skycoverage}
\end{figure}

The DEEP2 Galaxy Redshift Survey \citep{Newman13} used 90 nights with the DEIMOS spectrograph on the Keck-II telescope to measure the spectroscopic redshifts of $\approx 38,300$ galaxies in four near-contiguous regions of the sky, with a sky area of 2.8~square degrees.  The survey covered the wavelength range $6500$~\AA$-9100$~\AA\ with a high spectral resolution (R$=6000$), allowing a clean identification of the [O{\sc ii}]$\lambda$3727\AA\ doublet from the redshift range $z \approx 0.7-1.45$. The limiting apparent magnitude of the survey is R$_{\rm AB} = 24.1$ and the typical redshift accuracy, for objects with redshift quality $\geq 3$, is $\lesssim 62$~km~s$^{-1}$ \citep{Newman13}.

The DEEP2 Data Release 4 (DR4) catalogue \citep{Newman13} contains 27,966 galaxies with spectroscopic redshifts of quality~$\geq 3$ at $z=0.7-1.45$ in the four DEEP2 fields. 21,561  of these galaxies are located in DEEP2 fields 2, 3, and 4, which consist of seven sub-fields of size $\approx 28' \times 52'$. The full-width-at-half-maximum (FWHM) of the GMRT Band-4 receivers is $\approx 43'$ at 610~MHz, implying that each of the above seven sub-fields can be reasonably covered with a single GMRT Band-4 pointing (see Figure~\ref{fig:skycoverage}). Further, the \hii\ line from galaxies at $z=0.74-1.45$ is redshifted to the frequency range $\approx 580-820$~MHz, i.e. into the frequency coverage of the GMRT Band-4 receivers. DEEP2 fields 2, 3, and 4 are thus ideal targets for a GMRT \hii\ emission survey of galaxies at $z\approx 0.74-1.45$ \citep[e.g.][]{Kanekar16,Chowdhury20,Chowdhury21}. { We note that although DEEP2 Field~1, the Extended Groth Strip (EGS), has excellent multi-wavelength coverage \citep[e.g.][]{Davis07}, we did not observe it as the narrowness of the EGS \citep[$\approx 16'$; ][]{Newman13} implies that a significant fraction of the GMRT Band-4 primary beam would contain almost no galaxies with spectroscopic redshifts.}

The stellar masses of the DEEP2 galaxies were estimated via a relation between the (U$-$B) colour, the (B$-$V) colour and the ratio of the stellar mass to the B-band luminosity \citep{Weiner09}. This relation was calibrated at $z \approx 1$ using the subset of DEEP2 galaxies that have direct K-band estimates of the stellar mass. The scatter of individual stellar masses estimated using this relation is $\approx0.3$~dex \citep{Weiner09}.

\section{The Upgraded GMRT Observations}
\label{sec:obs}

\begin{table}
\begin{tabular}{|c|c|c|c|c|c|c|}
\hline

   DEEP2  & Right Ascension& Declination & GMRT & On-Source  & Number of  & $\sigma_\textrm{HI}$ \\
    subfield  & (J2000) & (J2000) & cycle & Time (hr)   & Galaxies & $\mu$Jy/Bm \\
    \hline \hline
     \multirow{2}{*}{21} & \multirow{2}{*}{$16$h$47$m$59.7$s} & \multirow{2}{*}{$+34\degree55'40.4''$} & 37 &  32.7 & 1,547 & 201  \\  
   \cline{4-7}
   & & & 38 &  17.9 & 1,545 & 313 \\ 
    \hline \hline
    
     \multirow{2}{*}{22} & \multirow{2}{*}{$16$h$51$m$28.9$s} & \multirow{2}{*}{$+34\degree56'58.9''$} & 37 &  22.8 & 1,623 & 274  \\  
   \cline{4-7}
   &  &  & 38 &  26.9 & 1,663 & 293 \\ 
    \hline \hline
    
    \multirow{3}{*}{31} &$23$h$26$m$52.8$s & $+00\degree08'25.7''$ & 35 &  14.3 & 1,390 & 328  \\ \cline{2-7} 
    & \multirow{2}{*}{$23$h$27$m$26.2$s } & \multirow{2}{*}{$+00\degree08'22.9''$} & 37 &  12.8 & 1,405 & 338 \\  \cline{4-7}
    & & & 38 &  33.9 & 1,226 & 232 \\ 
    \hline \hline
     \multirow{3}{*}{32} & \multirow{3}{*}{$23$h$29$m$49.9$s} & \multirow{3}{*}{$+00\degree12'12.7''$} & 35 &  14.5 & 1,403 & 304 \\  \cline{4-7}
    
    & & &  37 &  10.7 & 1,268 & 319 \\  \cline{4-7}
   & & & 38 &  26.7 & 1,431 & 288 \\ 
    \hline \hline
     \multirow{3}{*}{33} & \multirow{3}{*}{$23$h$32$m$58.7$s} & \multirow{3}{*}{$+00\degree08'22.7''$} &  35 &  15.3 & 1,227 & 293 \\  \cline{4-7}
    & & &   37 &  14.8 & 1,140 & 338 \\  \cline{4-7}
   & & & 38 &  33.9 & 1,062 & 195 \\ 
    \hline 
     \hline
     \multirow{3}{*}{41} & \multirow{3}{*}{$02$h$28$m$24.0$s} & \multirow{3}{*}{$+00\degree35'27.6''$} & 35 &  7.5 & 1,878 & 355  \\ \cline{4-7}
    & & & 37 & 31.4 & 1,899 & 286 \\  \cline{4-7}
   & & & 38 &  31.1 & 1,903 & 268 \\ 
    \hline
     \hline
     \multirow{3}{*}{42} & \multirow{3}{*}{$02$h$30$m$48.0$s} & \multirow{3}{*}{$+00\degree35'15.0''$} & 35 &  15.0 & 1,813 & 362 \\  \cline{4-7}
     & & &37 &  17.2 & 1,780 & 308 \\  \cline{4-7}
   & & & 38 &  28.0 & 1,790 & 291 \\ 
    \hline
     \hline
\hline
\end{tabular}
\caption{Summary of the GMRT observations and results.  The columns are : (1)~the DEEP2 subfield, (2,3)~the J2000 Right Ascension and Declination of the GMRT pointing centre, (4)~the observing cycle in which the observations were carried out, (5)~the on-source time, in hours, (6)~the final number of galaxies whose \hii\ subcubes were stacked, and (7)~the median RMS noise on the \hii\ subcubes at a velocity resolution of 30~\kmps\ and a spatial resolution of 90~kpc. See main text for discussion. Note that all observations had the same correlator setup, covering the frequency range $530-930$~MHz, with 8192 spectral channels.}
\label{table:obsSummary}
\end{table}






We used the upgraded GMRT \citep{Swarup91,Gupta17} Band-4 $550-850$~MHz receivers to observe DEEP2 fields 2, 3, and 4 for a total time of $\approx$510~hours, over three GMRT cycles between October~2018 and October~2020 (proposals 35\_087, 37\_063 and 38\_033; PI: Aditya Chowdhury). The total time was divided approximately equally between seven pointings in the three DEEP2 fields. This was motivated by (a)~minimizing the effect of cosmic variance on our measurements, by increasing the sky area, and (b)~reducing the risk of the root-mean-square (RMS) noise on the spectra of individual galaxies not decreasing $\propto 1/\sqrt{\Delta t}$ on a single field, due to systematic effects such as dynamic range issues. 

The DEEP2 sky area of $\approx2$~sq. degrees covered by the GMRT-CAT$z1$ survey (sky volume of $\approx10^7$ comoving Mpc$^3$) is sufficient to ensure that the effects of cosmic variance are negligible \citep{Newman02,Newman13}. \citet{Driver10} find that these effects would be $\lesssim10\%$ for a single contiguous survey volume of $\gtrsim10^7$~cMpc$^3$.  The total survey volume of $\approx10^7$~cMpc$^3$ of the GMRT-CAT$z1$ survey consists of three separate contiguous regions (DEEP2 fields 2, 3, and 4); the effects of cosmic variance are hence expected to be even lower than $\approx10\%$, by a factor of $\approx \sqrt{3}$ \citep{Driver10}.

The initial 90~hours of  observations were carried out between October 2018 and March 2019 (GMRT Cycle 35), with five pointings\footnote{The allocated time for the project in Cycle~35 was 100~hrs; we lost an entire 10~hour run on DEEP2 field 41 due to severe radio frequency interference.} on DEEP2 fields 3 and 4; the analysis of these data was presented in \citet{Chowdhury20}. Between October 2019 and March 2020 (GMRT Cycle 37), we used 170~hours of observations to observe all three DEEP2 fields with seven pointings. The remaining 250~hours were obtained between May 2020 and October 2020 (GMRT Cycle 38), again with seven pointings on the three DEEP2 fields. The total on-source observing time for each of the seven DEEP2 subfields was $\approx50-60$~hours. We note that the pointing centre on DEEP2 subfield~31 was shifted by $\approx8'$ between our observations of Cycle 35 and those of Cycles 37 and 38; this was done to reduce the deconvolution errors from bright radio-continuum sources at the edge of the pointing. Table~\ref{table:obsSummary} provides a summary of the 510~hours of observations, while Figure~\ref{fig:skycoverage} shows the different pointings on the three DEEP2 fields.

We used the GMRT Wideband Backend \citep{Reddy17} as the correlator, covering the frequency range $530-930$~MHz with a total bandwidth of 400~MHz, divided into 8,192 spectral channels. In each observing run, we used observations of one or more of the standard calibrators 3C48, 3C147, or 3C286 to calibrate the flux density scale, and observations of the nearby compact sources 0022+002, 0204+152, or 1609+266 to calibrate the antenna complex gains and antenna bandpass shapes. 

\section{Data Analysis, Sample Selection, and Statistical Tests}
\label{sec:analysis}

\subsection{The Data Analysis}


All data were analyzed in the Common Astronomy Software Applications package ({\sc casa} Version 5; \citealp{McMullin07}) following standard procedures. The same procedures, described in detail below, were used for the data from all observing cycles. We note that the analysis of the initial 90~hours of data from Cycle~35 was presented by \citet{Chowdhury20}. The new analysis described here contains a few minor improvements over that of \citet{Chowdhury20}.


The GMRT consists of 30 antennas, with 14 of these in a central 1~sq.~km. region (the ``central square'') and the remaining 16 arranged along three arms, forming a
``Y''-shape and providing baselines out to $\approx 25$~km. Terrestrial radio frequency interference (RFI) is typically stronger on the shorter central-square baselines, and decorrelates on the longer baselines. To reduce the effects of RFI, we entirely excluded the 91 baselines between pairs of central square antennas, and restricted our analysis to the remaining 344 baselines. The 91 central square baselines that were excluded have UV distances of $\approx 0.08 - 3.8$~k$\lambda$ at an observing frequency of $\approx 650$~MHz; these correspond to angular scales of $\approx 0.9'-43'$, far larger than the angular scales of interest in this work.

The data for each DEEP2 sub-field from the three GMRT observing cycles were analysed  independently. The result of the analysis is thus either two or three independent spectral cubes for each DEEP2 sub-field, one each for the data of each cycle. This was done to prevent systematic errors, such as those due to low-level RFI, imperfect deconvolution, etc, in the data of one GMRT cycle on a given field from affecting the overall data quality for that field. The independent spectral cubes allow us to separately examine the quality of the spectra from all DEEP2 galaxies from each GMRT cycle. For example, if the spectrum of a particular DEEP2 galaxy from Cycle~37 is affected by RFI, our approach ensures that the spectra of the galaxy from Cycles 35 and 38 can be used, if these are ``clean''.Conversely, if the data from all three cycles were combined to generate a single spectral cube for the pointing, we would have had to entirely exclude the DEEP2 galaxy in question from our sample.


For each observing run, after initially removing data from non-working antennas, the {\sc aoflagger} \citep{Offringa12} package was used to excise data affected by RFI. The antenna gains and bandpass shapes were determined using the data on the calibrator sources, and these solutions were applied to the target source visibilities. All antenna-based gain and bandpass calibration, as well as self-calibration, was performed using the calR\footnote{The package is publicly available at \url{https://github.com/chowdhuryaditya/calR} \citep{calR}} package \citep{Chowdhury20}, a collection of robust calibration routines within the {\sc CASA} framework.  Following this, all visibility data on each pointing from each cycle were combined together to produce a single multi-channel data set for each pointing. A standard iterative self-calibration procedure, along with further excision of data affected by low-level RFI, was then separately performed on the data of each pointing, again for each observing cycle. For each subfield, the data of the first observing cycle were calibrated using $3-4$ rounds of imaging and phase-only self-calibration, followed by $2-3$ rounds of imaging and amplitude-and-phase self-calibration. For observations of the same subfield in later cycles, the continuum image obtained from the first observing cycle was used as the initial self-calibration model (solving for both amplitude and phase),
followed by at least 2 rounds of imaging and amplitude-and-phase self-calibration. The imaging was done with the {\sc casa} task {\sc tclean}, with w-projection \citep{Cornwell08}, multi-frequency synthesis \citep[2nd-order expansion;][]{Rau11}, and Briggs weighting \citep{Briggs95}, with a robust parameter of $-0.5$.
For each field, the self-calibration procedure was continued until no improvement was seen in either the image or in the residuals after subtracting the image from the calibrated visibilities. 
At the end of the self-calibration procedure, the calibrated visibilities of each DEEP2 subfield of each cycle were separately imaged  with the task {\sc tclean}, again using w-projection, multi-frequency synthesis (2nd-order expansion), and Briggs weighting with a robust parameter of $0.0$.

We next used the final continuum images of each target field to subtract out all detected continuum emission from the self-calibrated spectral-line visibilities. For each field, the continuum subtraction was performed separately for the data in each observing cycle, using the continuum image of that cycle. We then used {\sc aoflagger} \citep{Offringa12} to perform one final round of RFI excision on the continuum-subtracted spectral-line visibilities.

\begin{figure}
    \centering
    \includegraphics[width=\linewidth]{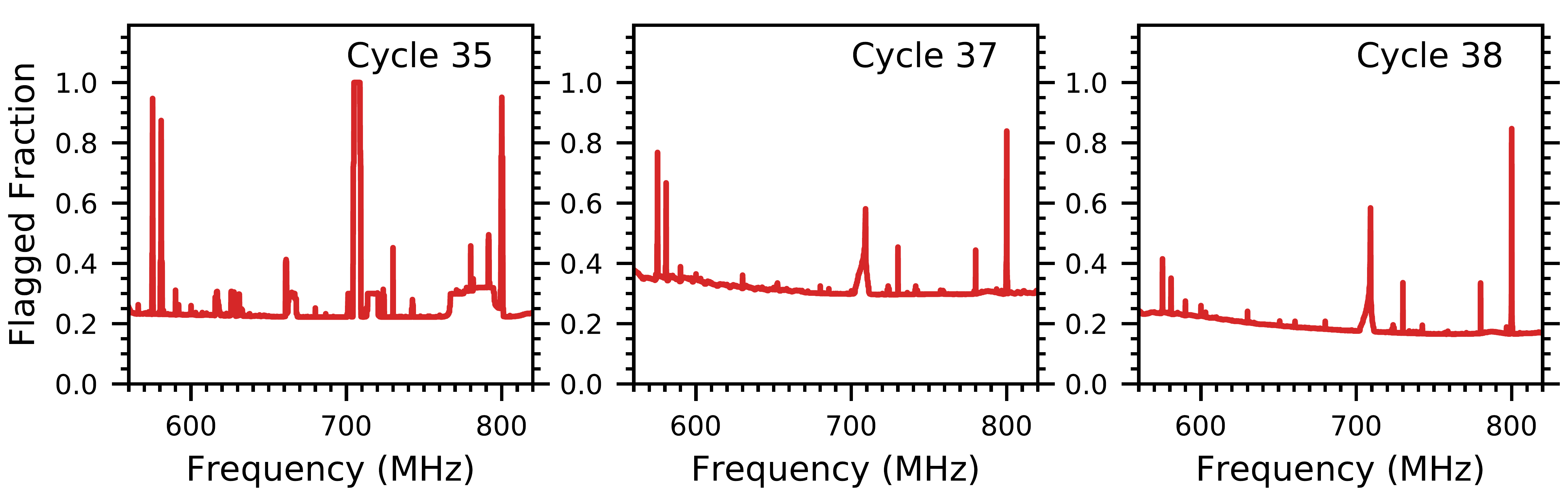}
    \caption{The fraction of data excised across the observing band due to time-variable issues (see main text for details), shown separately for the data obtained in each of the three GMRT cycles.}
    \label{fig:rfi}
\end{figure}

The fraction of data on each target field that was lost due to time-variable issues such as RFI, temporary problems with antennas, etc., but not including antennas that were not available throughout the observing run and the 91 baselines amongst the central-square antennas that were excised at the outset of the analysis, was $\approx 26\%$, $\approx 32\%$, and $\approx 19\%$ for Cycles~35, 37, and 38, respectively. Figure~\ref{fig:rfi} shows, separately for each observing cycle, the fraction of data excised due to such time-dependent issues as a function of observing frequency.

The {\sc casa} task {\sc tclean} was then used to make spectral cubes from the continuum-subtracted visibilities of each target field, and from each cycle; this yielded a total of 19 spectral cubes. The cubes were made in the barycentric frame, with w-projection and Briggs weighting with a robust parameter of $+1$. We experimented with various values of the robust parameter, and found Briggs weighting with a robust parameter of $+1$ to be the optimal choice; this results in reducing the effect of deconvolution errors around bright radio continuum sources, without causing a significant increase in the spectral RMS noise of the cubes. Each cube covers a sky area of $0.8^\circ \times 0.8^\circ$, larger than the FWHM of the GMRT primary beam at the lowest observing frequency of the band; this allowed us to obtain \hii\ spectra for all DEEP2 galaxies within the FWHM of the GMRT primary beam at the redshifted \hii\ line frequency of each galaxy. The channel width of the spectral cubes is 48.8~kHz, corresponding to a velocity resolution of $18-25$~\kmps\ across the frequency band. The synthesized beams of the cubes have FWHMs of  $4.0''-7.5''$, corresponding to spatial resolutions of $29-63$~kpc for the redshift range $z=0.74-1.45$.



\subsection{Radio continuum Images} 
	\begin{figure*}
		\centering
		\includegraphics[width=0.32\linewidth]{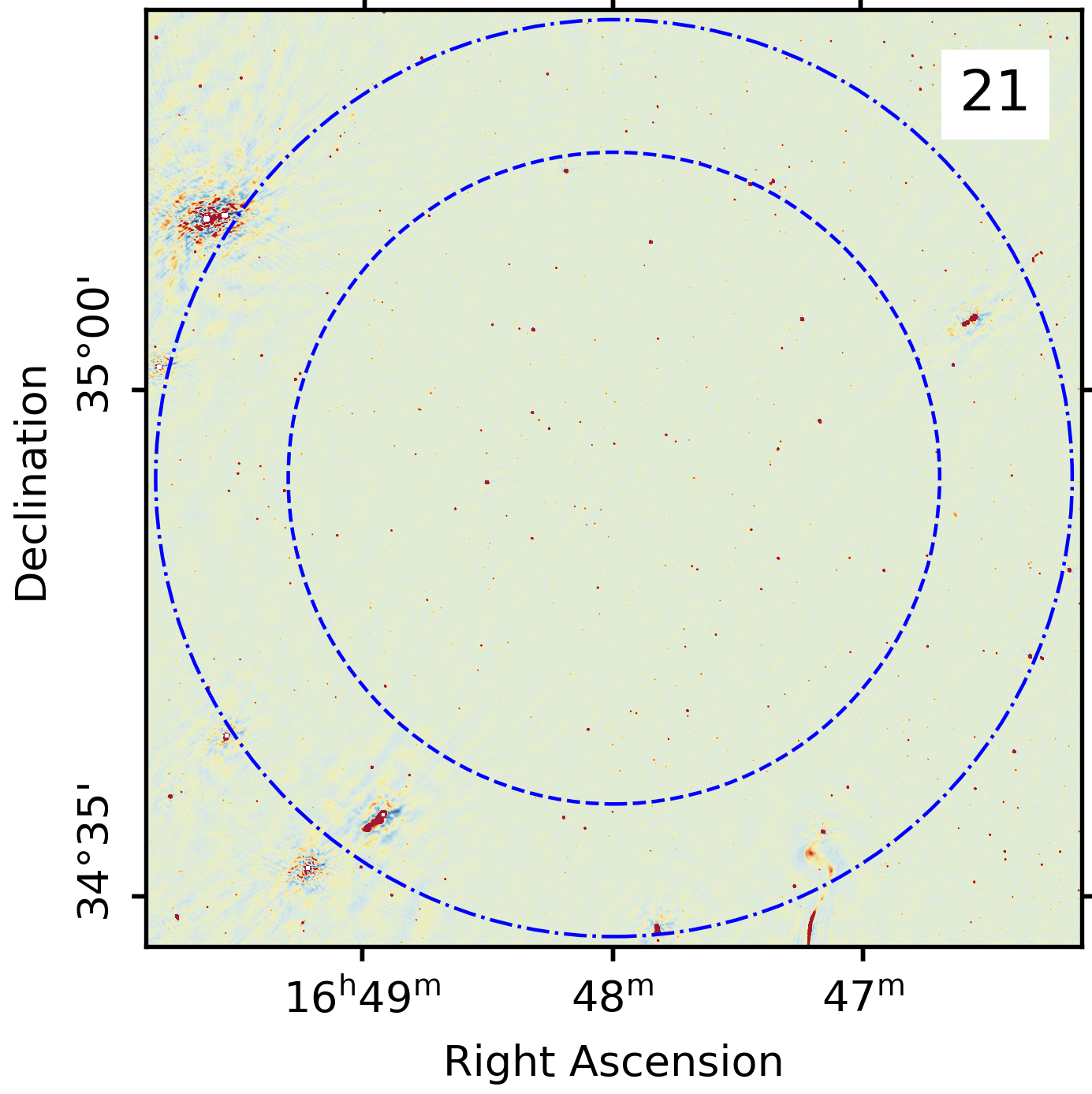}
		\includegraphics[width=0.32\linewidth]{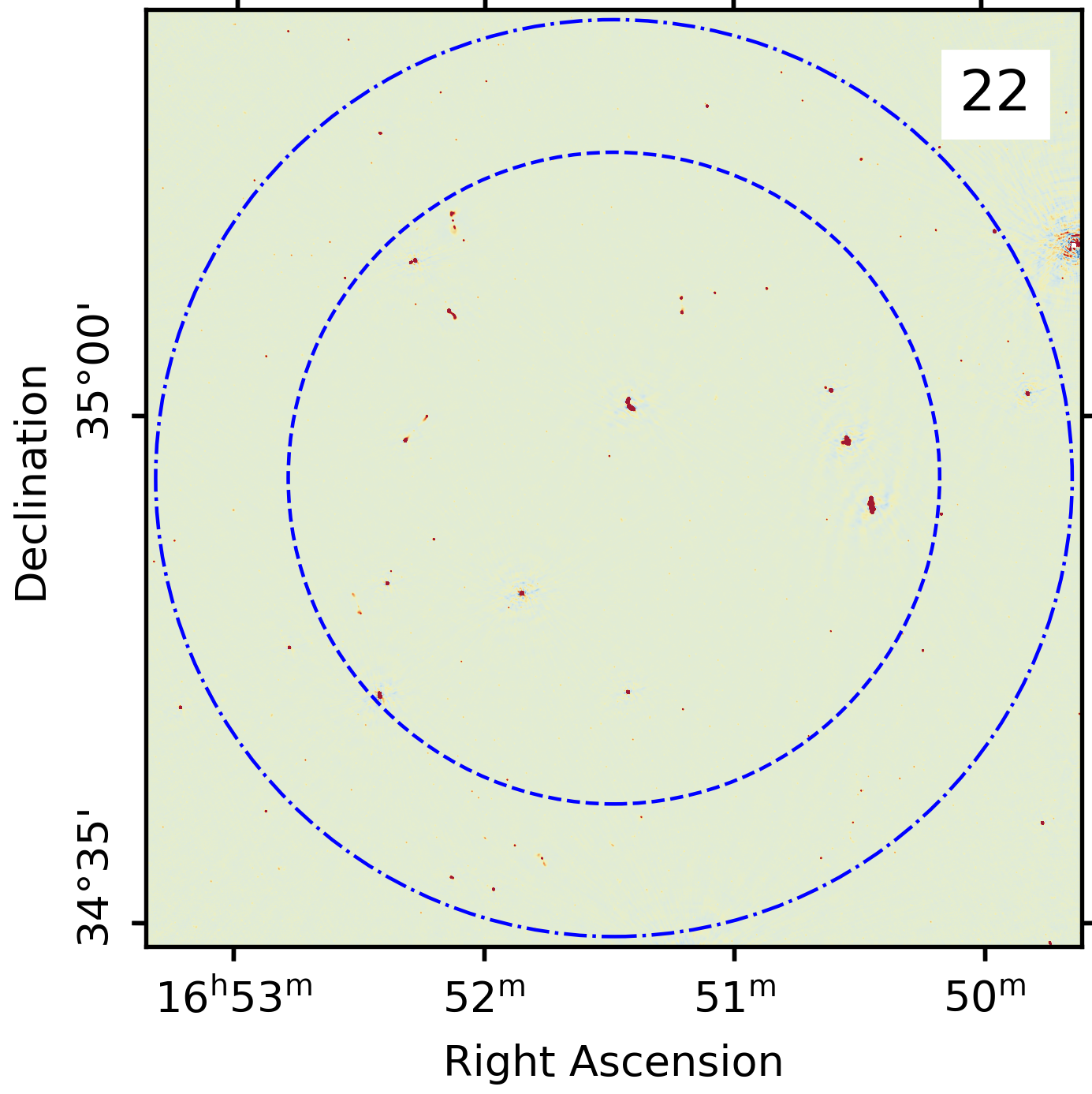}
		
		\includegraphics[width=0.32\linewidth]{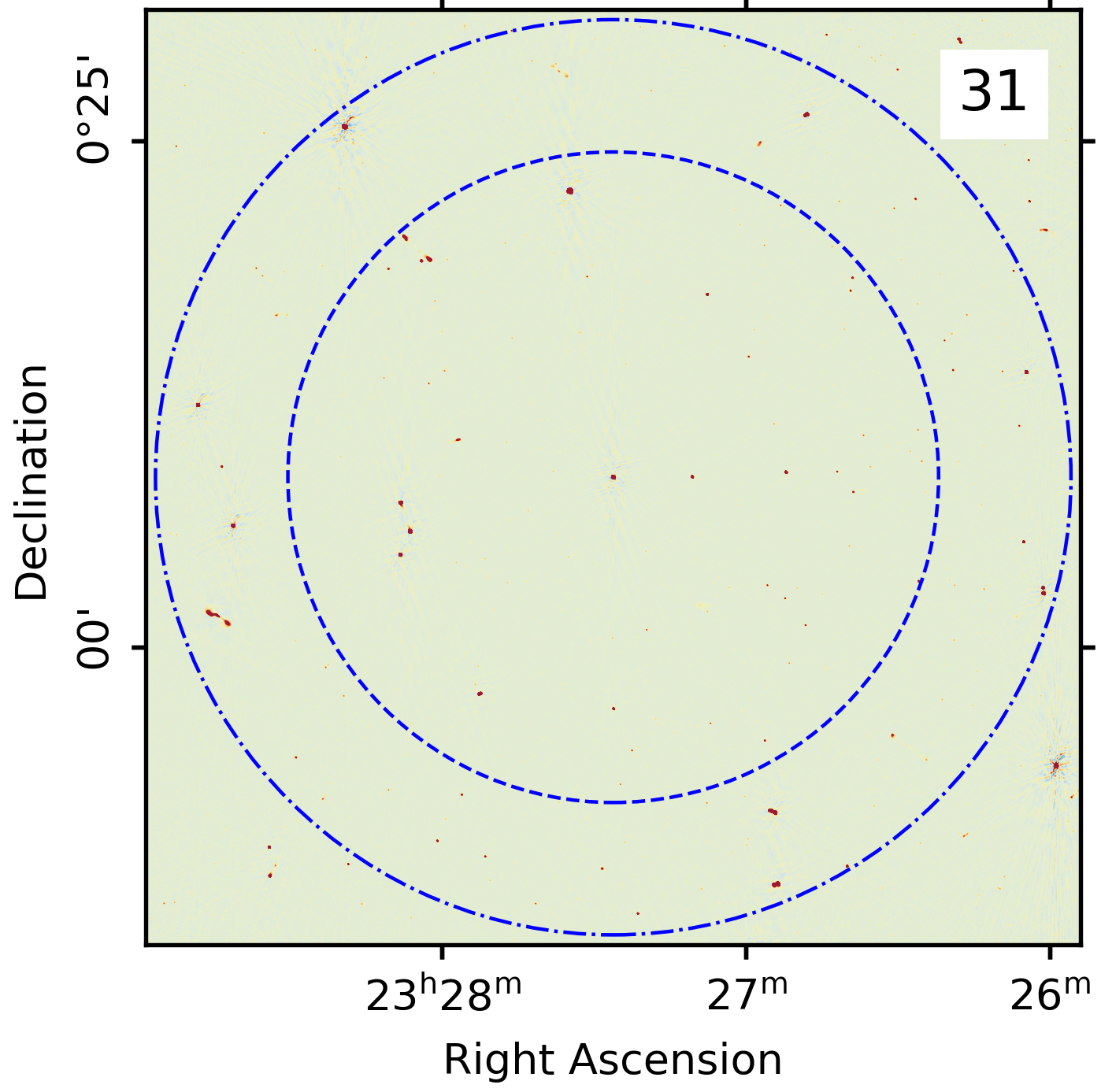}
		\includegraphics[width=0.32\linewidth]{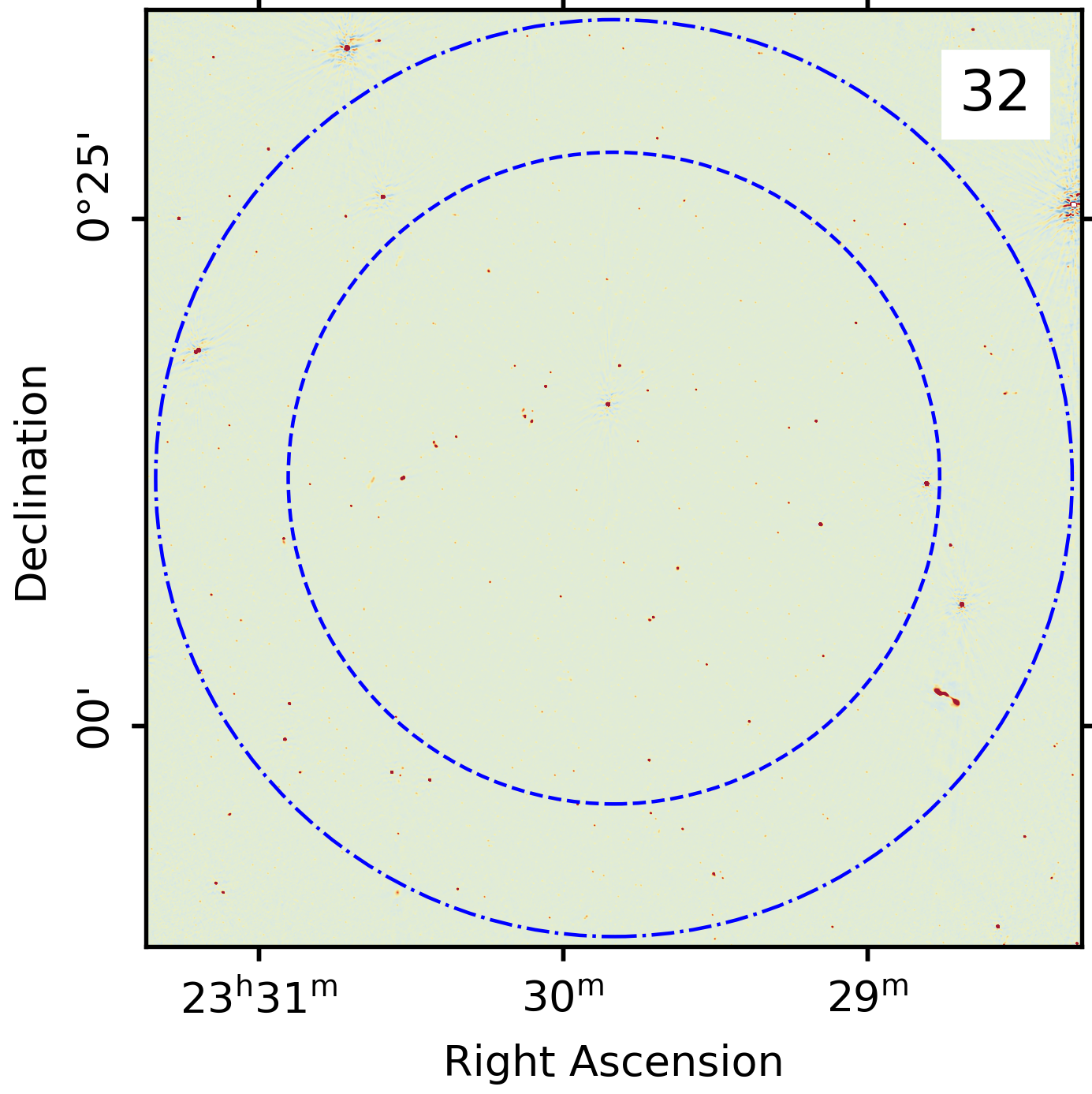}
		\includegraphics[width=0.32\linewidth]{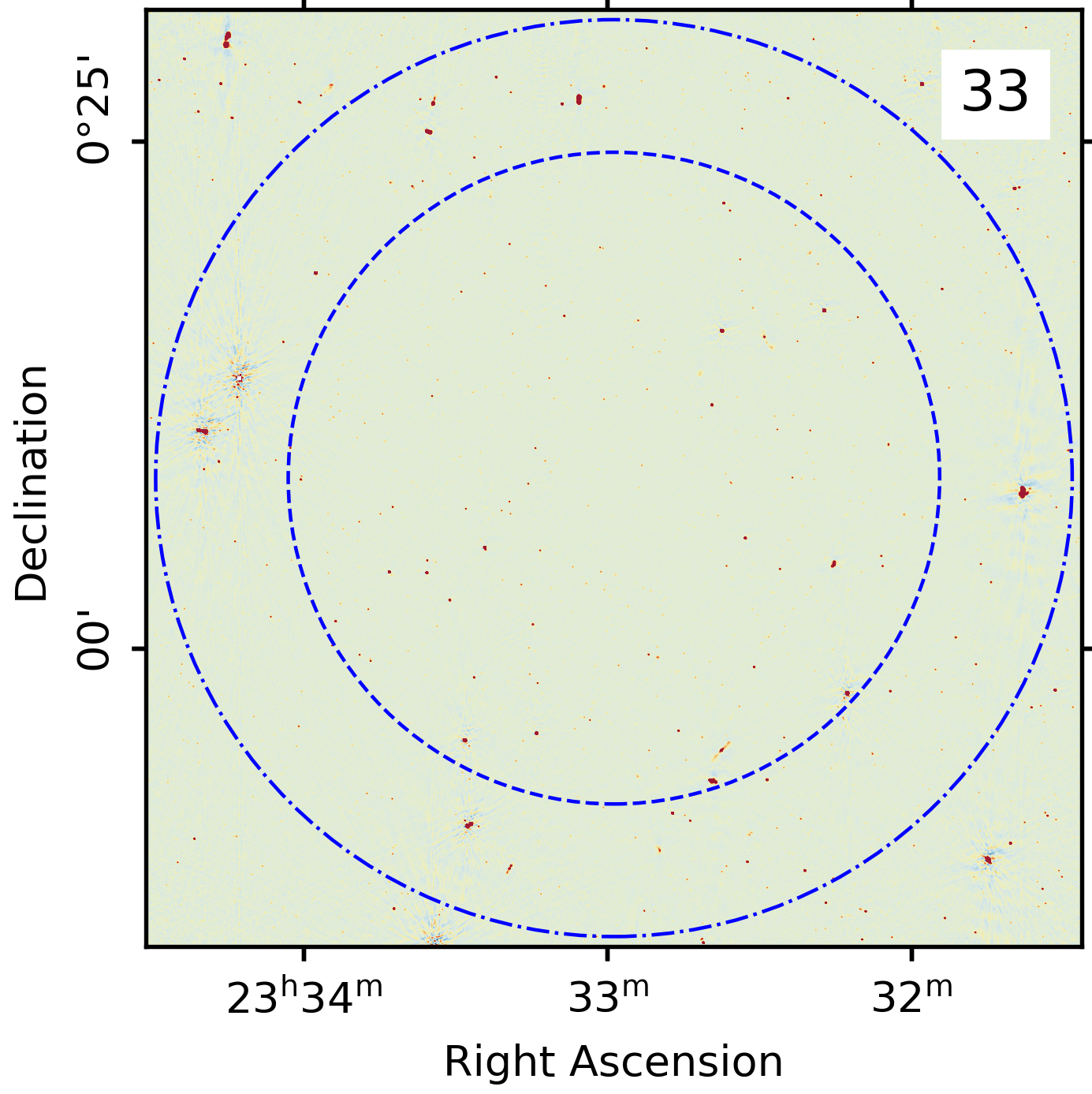}
		
		\includegraphics[width=0.32\linewidth]{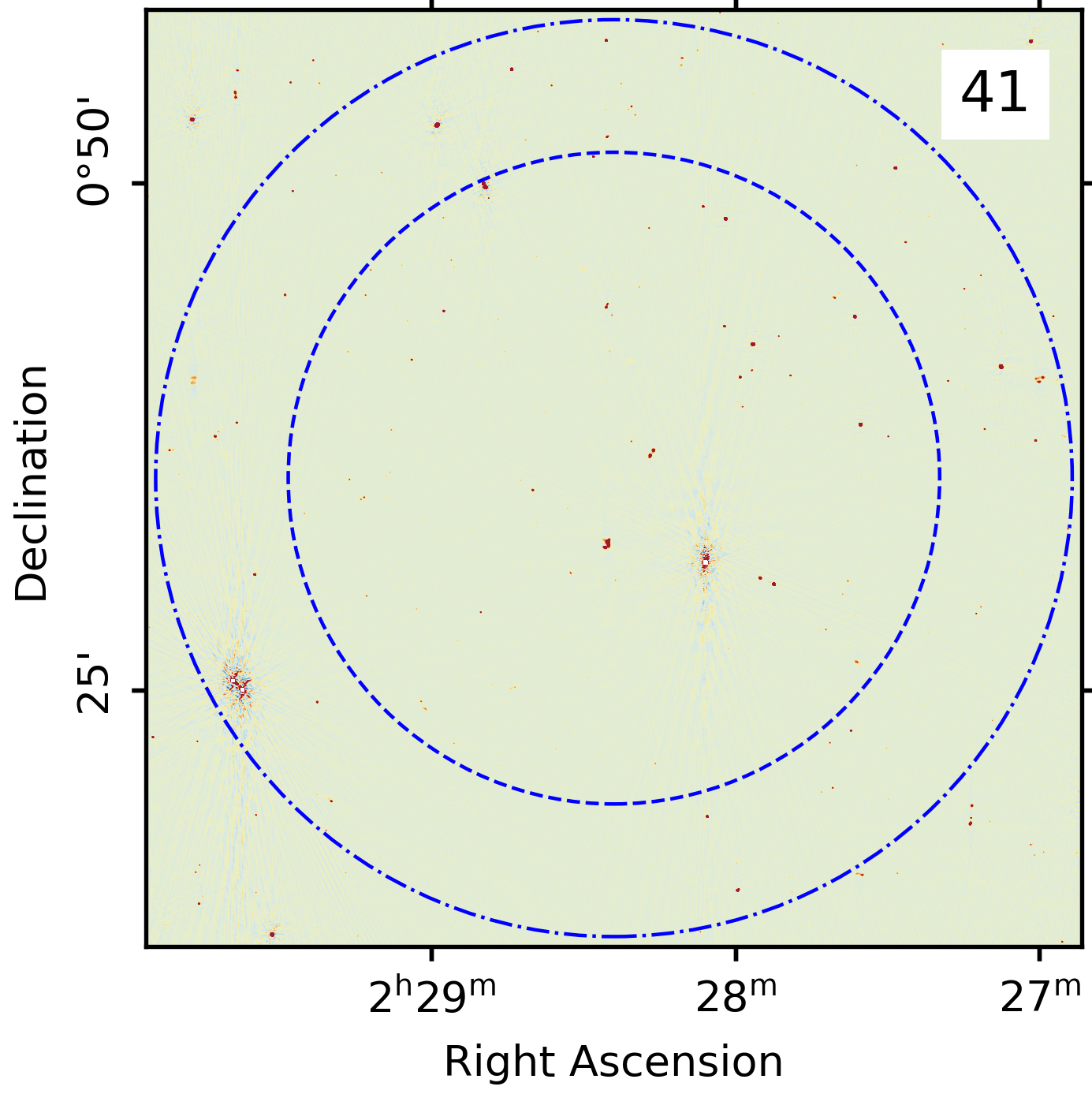}
		\includegraphics[width=0.32\linewidth]{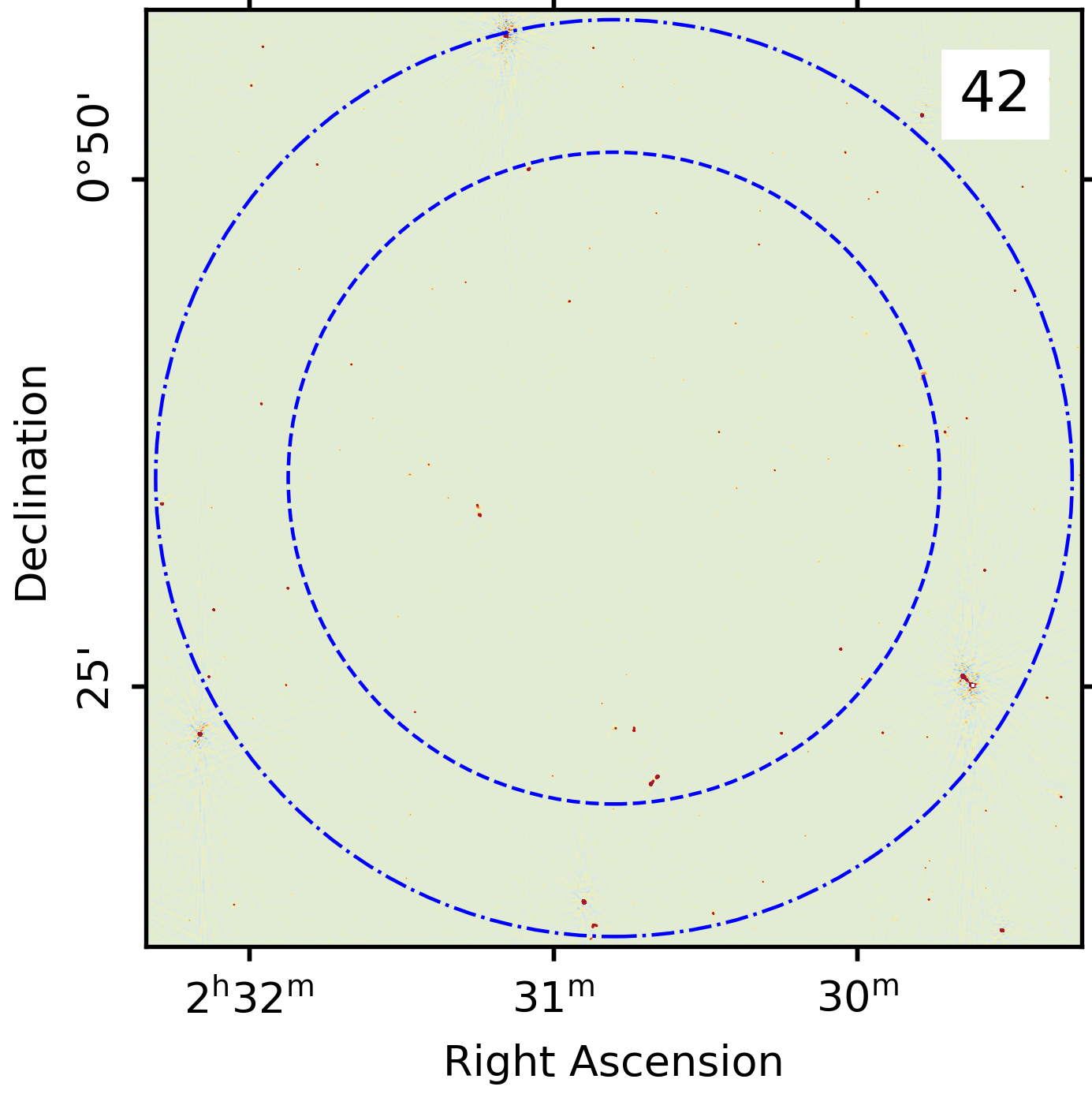}
		\caption{The GMRT 655~MHz continuum images of the seven DEEP2 sub-fields. The blue dashed-dotted and dotted circles in each panel show, respectively, the FWHM of the GMRT primary beam at $580$~MHz and $816$~MHz.}
		\label{fig:contimages}
	\end{figure*}
\begin{table}
\centering
\begin{tabular}{|c|c|c|c|}
\hline
\hline
   DEEP2  subfield & Synthesized Beam & $F_\textrm{max}$ & $\sigma_\textrm{cont}$   \\
                  & & mJy~Beam$^{-1}$ & $\mu$Jy~Beam$^{-1}$   \\
   \hline \hline
   21 & $3.9'' \times 3.1'' $ & 152.3 & 4.5 \\
   22 & $3.3'' \times 2.6'' $ & 158.5 &8.0  \\
   31 & $4.1'' \times 3.3'' $ & 82.4 & 6.2  \\
   32 & $4.1'' \times 3.3'' $ & 29.7 & 5.0  \\
   33 & $4.1'' \times 3.3'' $ & 34.4 & 4.6  \\
   41 & $3.7'' \times 2.9'' $ & 284.0 & 10.3 \\
   42 & $4.1'' \times 3.2'' $ & 75.6 & 5.9  \\
   \hline \hline
\end{tabular}
\caption{Properties of the GMRT 655~MHz continuum images of each DEEP2 subfield. The columns are (1)~the DEEP2 subfield, (2)~the FWHM of the synthesized beam, (3)~the peak flux density of the brightest source in the field, uncorrected for the primary beam response, and (4)~the RMS noise on the image, obtained by taking the median of the local RMS noise computed on regions of size $\approx 84''\times 84''$. Note that the continuum image of DEEP2 subfield 31 uses data from Cycles 37 and 38 only (see main text for details).}
\label{tab:cont}
\end{table}

The final 655~MHz radio continuum images of each DEEP2 subfield were made after combining all available data on each pointing. For each pointing, we first combined the calibrated visibilities from all cycles into a single data set, and then performed a few iterations of self-calibration on the combined data set. The self-calibrated visibilities of each subfield were then imaged, with w-projection, multi-frquency synthesis (2nd order expansion), and Briggs weighting with a robust parameter of $0.0$. For each pointing, we imaged a region of $\approx 1.5\degree \times 1.5 \degree$ around the pointing centre, extending well beyond the null of the GMRT primary beam at these frequencies. The  median of the RMS noise values in each continuum image is $\approx5-10~\mu$Jy~Beam$^{-1}$, while the FWHMs of the synthesized beams are $\approx 3''-4''$.  Table~\ref{tab:cont} lists the FWHMs of the synthesized beams and the RMS noise values of these final radio continuum images. The RMS noise values listed in the table were obtained by taking the median of the local RMS noise computed on regions of size $\approx 84''\times 84''$, without primary beam correction. The final continuum images of the seven DEEP2 subfields are shown in Figure~\ref{fig:contimages}. We note that the GMRT pointing on subfield 31 in Cycle~35 was slightly different from the pointing used in Cycles~37 and 38 (see Figure~\ref{fig:skycoverage}). The final radio continuum image of subfield~31 in Fig.~\ref{fig:contimages} is from the data obtained in Cycles~37 and 38.

\subsection{Sample Selection} 
\label{ssec:samplesel}
\begin{figure}
    \centering
    \includegraphics[width=0.7\linewidth]{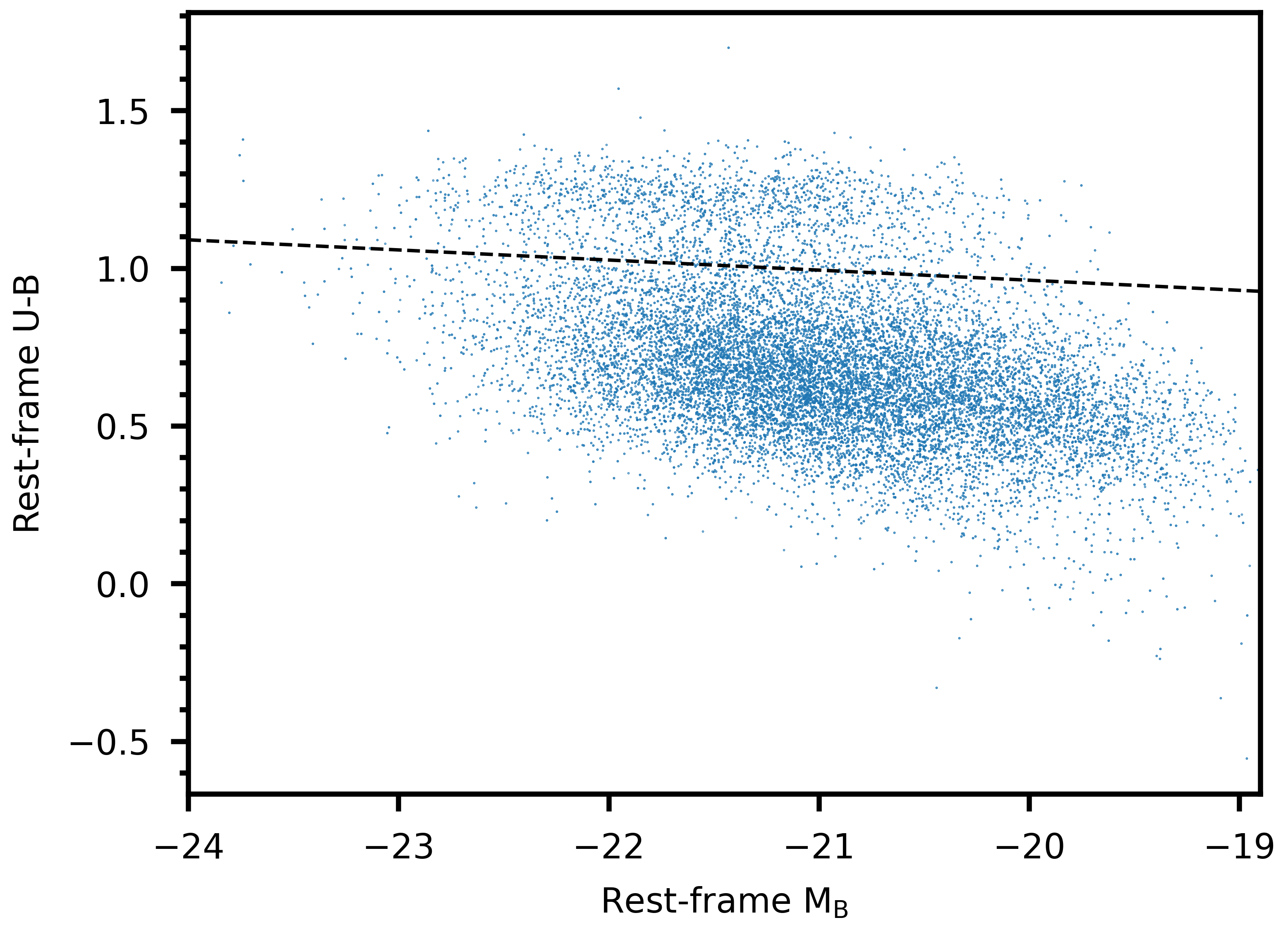}
    \caption{The rest-frame (U$-$B) vs rest-frame M$_\textrm{B}$ colour-magnitude diagram for the 16,250 DEEP2 galaxies covered by our observations. The blue points show the location of each DEEP2 galaxy in the diagram. The dashed line divides the sample into the red and blue galaxy populations \citep{Willmer06}. }
    \label{fig:colourmag}
\end{figure}
Our upgraded GMRT observations cover the redshifted \hii\ line for 16,250 DEEP2 galaxies with accurate redshifts \citep[redshift quality, Q$\ge$3 in the DEEP2 DR4 catalogue; ][]{Newman13} at $z=0.74-1.45$, lying within the FWHM of the GMRT primary beam at the redshifted \hii\ frequency of the galaxy. The FWHM of the GMRT primary beam is $\approx43'$ at 610~MHz, and scales with
frequency ($\nu$) as  FWHM~$\propto 1/\nu$. The \hii\ lines of the 16,250 galaxies at $z=0.74-1.45$ are redshifted to frequencies $\approx 580-820$~MHz, where the Band-4 receivers have their highest sensitivity.

The DEEP2 survey is magnitude-limited in the R-band (R~$< 24.1$) and this preferentially picks out blue objects at $z>0.7$ \citep{Willmer06,Newman13}. Indeed, only 2,222 out of the 16,250 DEEP2 objects covered by our observations are part of the ``red cloud" in the color-magnitude diagram \citep{Willmer06}. Figure~\ref{fig:colourmag} shows the distribution of 16,250 DEEP2 galaxies in the rest-frame (U$-$B) vs rest-frame M$_\textrm{B}$ colour-magnitude diagram. We ensure the homogeneity of our sample by restricting our study to the 14,028 blue objects at $z=0.74-1.45$ that are covered by our observations. These 14,028 galaxies lie below the green valley in the rest-frame (U$-$B) vs rest-frame M$_\textrm{B}$ colour-magnitude diagram (see Figure~\ref{fig:colourmag}), and were selected using the criterion  C$\le0$, where the colour ${\rm C=U-B}+0.032 \times ({\rm M_B}+21.63)-1.014$ \citep{Willmer06}.

Some of the above 14,028 blue DEEP2 galaxies may host AGNs, which could affect the gas properties of the galaxy. We hence used our radio-continuum images of the DEEP2 subfields to exclude all detected radio AGNs from our sample. Radio-continuum studies show a bimodality in the distribution of rest-frame 1.4-GHz radio luminosity (L$_{1.4\ \textrm{GHz}}$), with objects having L$_{1.4\ \textrm{GHz}}>2\times10^{23}$~W~Hz$^{-1}$ being predominantly radio-bright AGNs, and objects with L$_{1.4\ \textrm{GHz}}<2\times10^{23}$~W~Hz$^{-1}$ being predominantly  star-forming galaxies \citep{Condon02}. We use this criterion to exclude the 882 DEEP2 objects which are detected at $\geq 4\sigma$ statistical significance in our radio-continuum images, with L$_{1.4\ \textrm{GHz}}>2\times10^{23}$~W~Hz$^{-1}$. After excluding these 882 radio-bright AGNs, our sample contains 13,146 blue star-forming galaxies at $z=0.74-1.45$.

Finally, we excluded from the sample the 487 galaxies that have stellar masses $\Ms<10^{9} \ \Msun$. This was done in order to ensure that our results can be directly compared with results for the xGASS survey \citep{Catinella18}, which has measured the \hi\ properties of galaxies at $z \approx 0$ with $\Ms>10^{9} \ \Msun$. We note that only $\approx4\%$ of our galaxies have $\Ms<10^{9} \ \Msun$ and their exclusion from the sample does not affect the results presented in this work (within the statistical uncertainties). After excluding the 487 galaxies with $\Ms<10^{9} \ \Msun$, our sample contains 12,659 blue star-forming galaxies at $z=0.74-1.45$. 

\subsection{\hii\ spectral subcubes}
\label{ssec:hisubcubes}

We obtain multiple \hii\ spectra, with uncorrelated statistical noise, for nearly every blue star-forming DEEP2 galaxy in our sample\footnote{Henceforth, for brevity, we will refer to \hii\ spectra with uncorrelated statistical noise as ``independent'' \hii\ spectra.}.
Specifically, for each DEEP2 galaxy, we obtain 2 or 3 independent \hii\ spectra from the observations in the different GMRT cycles. We also obtain additional independent \hii\ spectra for the galaxies that lie in the overlap regions of our pointings on the DEEP2 subfields. A total of 33,640 independent \hii\ spectra were obtained for the 12,659 galaxies in our sample, with up to six independent \hii\ spectra per DEEP2 galaxy. 



For each DEEP2 galaxy, we extracted an \hii\ subcube centred on the galaxy from each spectral cube that covered its \hii\ line. Each subcube covers an angular extent of $\pm76.8''$ around the galaxy location, and a velocity range of $\pm1500$~\kmps\ around its redshifted \hii\ line frequency. We convolved each of the 33,640 independent \hii\ subcubes with Gaussian beams such that the FWHMs of the final beams of each convolved subcube are identical. 

Next, the spatial extent of the \hii\ emission from galaxies at these redshifts is not known. We hence initially convolved each subcube to a range of spatial resolutions, 60~kpc, 70~kpc, 80~kpc, 90~kpc, 100~kpc, 110~kpc, 120~kpc, 150~kpc, and 200~kpc. The analysis of the subcubes was carried out at each of the above resolutions. Our final choice of the spatial resolution is described in Section~\ref{sec:massresolution}
 below.

During the process of convolution, care was taken to normalize the convolved subcubes such that the peak of the convolved point-spread function (using the same kernel) is unity \citep{Chowdhury20}. Any spectral channel where the intrinsic beam has a FWHM~$>60$~kpc was excised at this stage. Next, we regridded each subcube to a uniform spatial and spectral grid,  with $10$~kpc spatial pixels covering $\pm 500$~kpc around the position of the DEEP2 galaxy, and $30$~\kmps\ velocity channels covering $\pm1500$~\kmps\ around its redshifted \hii\ line frequency.  Finally, we fitted a second-order polynomial to the spectrum at each spatial pixel of the subcubes, and subtracted this out. This was done to remove any residual spectral baselines due to deconvolution errors from bright radio-continuum sources.

\subsection{Statistical tests for systematic effects in the spectral cubes}
\label{ssec:spectraflag}

 We carried out a series of tests on the subcubes, aiming to remove any subcubes affected by non-Gaussian systematic effects. This was done to ensure that the noise properties of the stacked \hii\ subcubes are not limited by such systematics. Initially, we discarded any \hii\ subcube which has $>15\%$ of its spectral channels excised due to RFI; this resulted in the excision of 3,721 \hii\ subcubes, i.e. $\approx 11$\% of the full sample.

 Next, we tested whether the noise properties of the \hii\ subcubes are consistent with their arising from a Gaussian distribution. The tests were carried out on the subcubes with a spatial resolution of 60~kpc; performing these tests at coarser spatial resolutions yielded similar results. We excluded spectra based on the following criteria :
\begin{enumerate}
        \item Any \hii\ subcube having a spectral feature of $\geq 6.0\sigma$ significance, at either the native velocity resolution ($30$~\kmps), or after smoothing to resolutions of $60$~\kmps\ and $120$~\kmps, was rejected.

     \item Each \hii\ subcube was tested for the presence of correlations between neighboring spectral channels (e.g., due to a residual spectral baseline) by examining the decrease in the RMS noise after smoothing to coarser velocity 
    resolutions. Specifically, each \hii\ subcube was smoothed by a factor of 10 to a spectral resolution of 300~\kmps; any subcube whose RMS noise was found to decrease by a factor $>0.47$ after the smoothing was excluded from the sample.
\end{enumerate}

These tests resulted in our excluding 926 \hii\ subcubes, i.e. $\approx3\%$ of the remaining 29,919 \hii\ subcubes. Our final sample, after excising 4,647 subcubes in all, contains a total of 28,993 independent \hii\ subcubes for 11,419 blue star-forming galaxies at $z=0.74-1.45$.  Table~\ref{table:obsSummary} lists  the final number of galaxies (i.e. \hii\ subcubes) obtained from the observations of each subfield in
each GMRT cycle.  We emphasize that the results of this paper are not sensitive to the exact choice of the thresholds used in the above tests.

\subsection{The RMS noise on the individual \hii\ }
\label{sec:rms-noise}

The presence of systematic effects in the data, such as deconvolution errors, residual low-level RFI, etc., can limit the final spectral RMS noise obtained on the \hii\ spectra. To test for such effects, we computed the RMS noise on our final sample of 28,993 independent \hii\ subcubes, at a spatial resolution of 90~kpc, and compared this to the spectral RMS noise expected for upgraded GMRT observations. For a given observation of a DEEP2 subfield, we compute the expected spectral RMS noise taking into account (1)~the sensitivity of the upgraded GMRT Band-4 receivers (as measured by the observatory), (2)~the total on-source time obtained on the DEEP2 subfield in that GMRT cycle, (3)~the fraction of data excised due to RFI, non-working antennas, power failures, etc., (4)~the effect of spectral smoothing  to obtain a velocity resolution of 30~\kmps, and (5)~the effect of spatial smoothing to obtain a resolution of 90~kpc.

Figures~\ref{fig:rmsfield2}, \ref{fig:rmsfield3}, and \ref{fig:rmsfield4} in Appendix~\ref{appndx:rmsnoise} compare the spectral RMS noise obtained on the \hii\ subcubes of the DEEP2 galaxies with the expected RMS noise for each observation. We find that the actual RMS noise obtained on our \hii\ subcubes is within $\approx 10\%$ of the RMS noise predicted using the GMRT sensitivity curve for the Band-4 receivers. We note that the GMRT sensitivity curve is itself only accurate to $\approx 10\%$, while the typical flux-scale uncertainty for upgraded GMRT observations (such as the ones presented here) is also $\approx 10\%$. We thus find no evidence for systematic effects that might affect the spectral RMS noise on our sample of 28,993 independent \hii\ subcubes. We also repeated the above analysis at other spatial resolutions ($> 60$~kpc) and find similar results. Overall, we find no evidence suggesting that the final RMS noise obtained on our \hii\ subcubes might have been limited by systematic effects.


\section{The Stacking Analysis}
\label{sec:stacking}

\subsection{Stacking the \hii\ Emission}
\label{sec:stackingHI}
 For each \hii\ subcube, we first corrected the measured flux density to take into account the position of the DEEP2 galaxy in the GMRT primary beam, and then converted  the flux density to the corresponding \hii\ line luminosity density, ${\rm L_{H\textsc{i}}}$ (in units of Jy Mpc$^2$), using the relation ${\rm L_{H\textsc{i}}}=4\pi \ 	{\rm S_{H\textsc{i}}} \ {\rm D_L}^2/(1+z)$, where $\rm S_{H\textsc{i}}$ is the \hii\ line flux density (in units of Jy), and ${\rm D_L}$ is the luminosity distance of the galaxy (in units of Mpc). The stacked \hii\ spectral cube for the DEEP2 galaxies was obtained by taking an average, across all \hii\ subcubes of the final sample, of the \hii\ line luminosity densities in the corresponding spatial pixels and velocity channels of the individual subcubes. We then fitted a second-order spectral baseline to the spectrum of each spatial pixel of the stacked cube, excluding the central $\pm250$~\kmps\ velocity range, and subtracted out this baseline from each spatial pixel to obtain the final stacked \hii\ spectral cube.

We determined the RMS noise on the stacked  cube by using Monte Carlo simulations in which we shifted the central velocity of each DEEP2 galaxy in the range $\pm1500$~\kmps, and then stacked the velocity-shifted \hii\ subcubes. Spectral channels that were shifted outside the $\pm 1500$~\kmps\ velocity range were wrapped around to the other side of the spectrum, before the stacking. We repeated the above procedure to obtain 10$^4$ realizations of the stacked \hii\ subcube. For each spatial and velocity pixel of the stacked \hii\ subcube, we computed the RMS of the \hii\ luminosities across the $10^4$ realizations to obtain an estimate of the RMS noise on the pixel.



Finally, we used a boxcar kernel to smooth the stacked \hii\ subcubes, including those derived via the Monte Carlo simulations, to a velocity resolution of 90~\kmps. {This was done in order to increase the signal-to-noise ratio (SNR) per channel, to clearly identify the velocity range with \hii\ emission.} We used these subcubes to estimate the average \hi\ mass of the DEEP2 galaxies, and the error on the average \hi\ mass. {We note that our results are not sensitive to the exact choice of the final velocity resolution of the stacked \hii\ cubes. The average \hi\ mass derived from the stacked \hii\ spectra at other resolutions (e.g. 60~\kmps, 120~\kmps, etc) are consistent with those reported in this paper, for a velocity resolution of $90$~\kmps.}

We also computed the errors on the stacked \hii\ subcube via two other methods: (i)~bootstrap resampling with replacement, and (ii)~computation of the RMS noise from the image plane of the stacked subcube. The errors derived via these methods are very similar to those derived via our Monte Carlo approach.

 The average \hi\ mass of our sample of DEEP2 galaxies was obtained as follows:  (i)~the central velocity channels of the final stacked cube were integrated to produce an image of the \hii\ emission, (ii)~the stacked \hii\ spectrum was obtained by taking a cut through the location of the peak luminosity density in this \hii\ image, (iii)~a contiguous range of central velocity channels in the stacked \hii\ spectrum, each containing emission at $\geq 1.5\sigma$ statistical significance, was selected, (iv)~the signal in these channels was integrated to obtain the average velocity-integrated \hii\ line luminosity density ($\int {\rm L_{H\textsc{i}} \ dV}$), in units of ${\rm Jy~Mpc^2}$~\kmps, and (v)~the average \hi\ mass of the sample was obtained via the expression $\MHI=[1.86 \times 10^4 \times \int {\rm L_{H\textsc{i}} \  dV}]~\Msun$.
 
 We note that the average \hi\ masses obtained by integrating the stacked \hii\ spectra over a wide velocity range of $\pm250$~\kmps\ are consistent with those obtained from the above approach, integrating over a contiguous range of central velocity channels with emission at $\geq 1.5\sigma$ statistical significance.

\subsection{Stacking the rest-frame 1.4~GHz continuum emission}
\label{sec:stackingCont}

Measurements of the rest-frame 1.4~GHz radio luminosity of a star-forming galaxy can be used to infer its SFR via the known correlation between the radio and the far-infrared (FIR) luminosities \citep{Condon92,Yun01}. We use our radio-continuum images of the DEEP2 subfields, along with the FIR-radio correlation, to estimate the average SFR of our galaxies  \citep[e.g.][]{White07,Bera18,Leslie20}.

  The synthesized beams of our radio-continuum images correspond to a physical scale of $\approx 24-35$~kpc over the redshift range $z=0.74-1.45$. We extracted $\approx 84''\times84''$ sub-images around each DEEP2 galaxy of our sample and convolved each sub-image with a Gaussian kernel such that the final synthesized beam has an FWHM of 40~kpc (again normalizing the sub-images such that the peak of the convolved point-spread function is unity; \citealp{Chowdhury20}). We then regridded the sub-images to a uniform grid with 5.2~kpc pixels and extending over $\pm 260$~kpc. Next, for each DEEP2 galaxy, we converted the observed flux density in each pixel to the rest-frame 1.4~GHz luminosity at the galaxy redshift, assuming a spectral index of $\alpha=-0.8$ \citep{Condon92}, with $S_{\nu} \propto \nu^{\alpha}$. We note that the central frequency of our radio-continuum images corresponds to rest-frame frequencies of $\approx1.14-1.61$~GHz for the galaxies of our sample, quite close to 1.4~GHz; our results are hence insensitive to the exact choice of $\alpha$. A median stacking approach  was then used to estimate the average rest-frame 1.4~GHz luminosity of our sample \citep{White07}. In this approach, we compute the median of the 1.4~GHz luminosities of each spatial pixel, across the sample of galaxies. The median rest-frame 1.4~GHz luminosity is then converted to a median SFR using the relation SFR~$(\Msun/\textrm{yr}) = (3.7 \pm 1.1) \times 10^{-22} \times {\rm L_{1.4 GHz} \ (W/Hz)}$ \citep[][after scaling to a Chabrier IMF]{Yun01}.

In order to test for systematic effects and to compute the RMS noise on our stacked continuum image, we repeated the above procedure for sub-images at locations offset by 100$''$ from the DEEP2 galaxies. The stack at the offset locations show that the mean of the distribution of luminosity-density values is slightly shifted to negative values; in flux-density units, the offset is $\approx -0.4 \ \mu$Jy, more than an order of magnitude lower than the RMS noise on each of our radio-continuum images. This weak bias in the mean of the distribution is likely to be due to negative sidelobes of very faint sources in our continuum maps, i.e. those with flux densities comparable to or lower than our cleaning thresholds. For the rest-frame 1.4~GHz continuum stack of a given subsample of galaxies, we correct for this effect by (i)~computing the mean luminosity density of the stack at locations offset by 100$''$ from the target galaxies, and (ii)~subtracting this mean luminosity density from all pixels of the stacks at both the galaxy locations and the offset locations. Effectively, this amounts to a small zero-point correction in our continuum stacked images; the correction is at the level of $\approx4-5\%$ of the detected 1.4~GHz luminosity in the stacked radio-continuum images.

We note that uncertainties in the flux density scale of our radio-continuum images would affect our radio-derived SFR estimates. These systematic uncertainties are typically  $\lesssim 10\%$ for the GMRT, for our calibration procedure. We have very conservatively treated this as a $1\sigma$ error on the flux density estimates. Thus, the errors on our SFR values include both a $10\%$ systematic uncertainty and the $1\sigma$ statistical uncertainty.

In passing, we note that we are unlikely to miss any radio-continuum emission from the star-forming regions of the galaxies of our sample because the 40-kpc beam of the stacked radio-continuum images is much larger than the observed sizes of star-forming regions in galaxies at these redshifts \citep{Trujillo04}. Indeed, the FWHM of the R-band emission is less than 10~kpc for $\approx98\%$ of the 11,419 DEEP2 galaxies in our sample \citep{Coil04}.

\section{The Sample of 11,419 Blue, Star-forming Galaxies}
\label{sec:sample}

	Our main sample consists of 11,419 blue star-forming galaxies at $z=0.74-1.45$ in the seven uGMRT pointings on DEEP2 fields 2, 3 and 4, after excluding AGNs, red galaxies, galaxies with stellar masses $\Ms < 10^9 \ \Msun$ (see Section~\ref{ssec:samplesel}), and galaxies whose \hii\ subcubes were affected by discernible systematic effects (see Section~\ref{ssec:spectraflag}). Our upgraded GMRT observations provide up to six independent measurements of the \hii\ spectrum of each galaxy of our sample (see Section~\ref{ssec:hisubcubes}). 
	Approximately $57\%$ of the galaxies  have $\geq 3$ independent measurements of their \hii\ spectra, with at least one from each GMRT cycle. In all the analysis presented in this paper, we have treated the independent \hii\ spectra as arising from separate ``objects''. This effectively implies that, in computing the average quantities of the sample, each of the 11,419 sample galaxies has a weight proportional to the number of independent measurements of its \hii\ spectrum. 

\subsection{Redshifts and Stellar Masses}
\label{ssec:z-mstar}


\begin{figure}
    \centering
    \includegraphics[width=\linewidth]{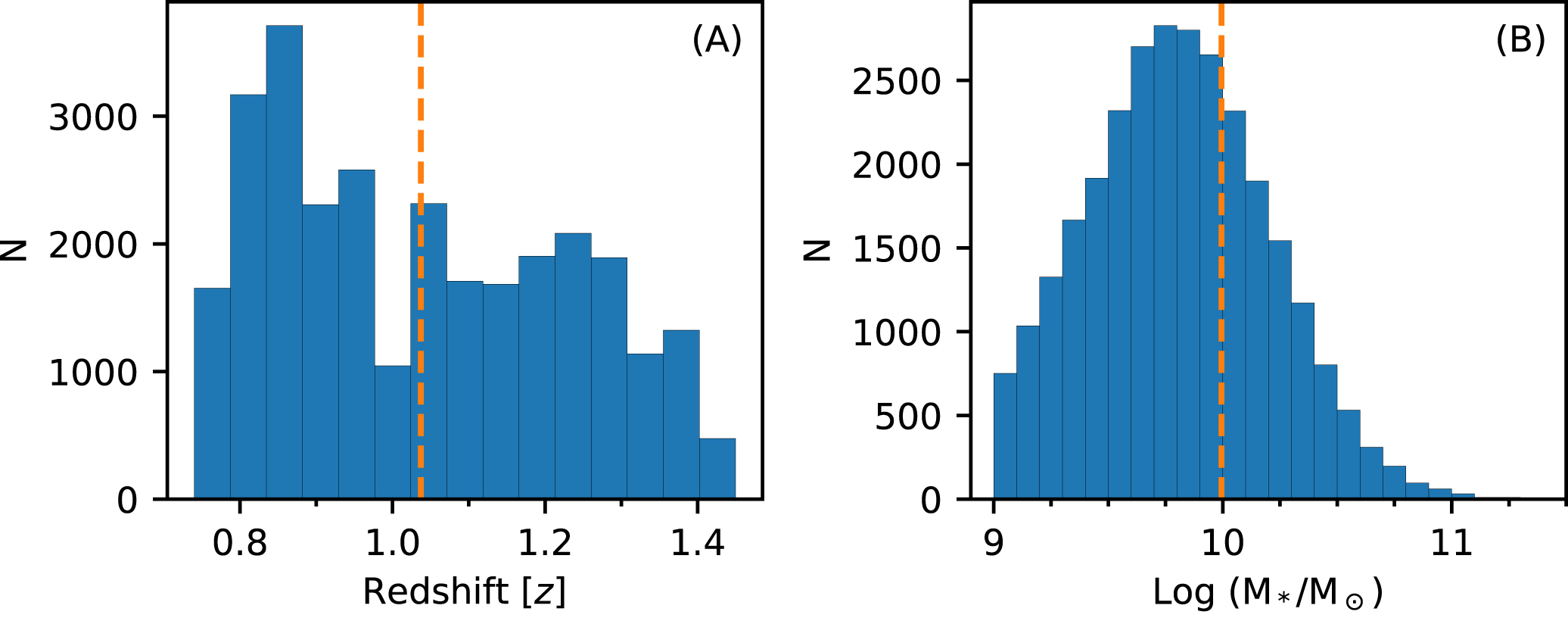}
    \caption{[A]~The redshift distribution of our main sample of 11,419 galaxies. The histogram shows the number of independent \hii\ spectra as a function of redshift. [B]~The stellar-mass distribution of the sample. The histogram shows the number of independent \hii\ spectra in each logarithmic stellar-mass bin. The dashed orange lines in Panel~[A] and Panel~[B] show, respectively, the average redshift and the average stellar mass of the sample. 
    }
    \label{fig:distall}
\end{figure}

The redshift distribution of the 11,419 galaxies of our sample, after accounting for the number of independent \hii\ spectra per galaxy, is shown in Figure~\ref{fig:distall}[A]. The sample spans the redshift range $z=0.74-1.45$, with a mean redshift $\langle z \rangle=1.04$.

Our sample of 11,419 galaxies has stellar masses in the range $\Ms=10^{9}-10^{11.4}\Msun$. The stellar mass distribution of the 11,419 galaxies, after factoring in the number of independent \hii\ spectra per galaxy, is shown in Figure~\ref{fig:distall}[B]. The mean stellar mass of the sample is $\langle \Ms \rangle=9.9\times10^{9}~\Msun$. 

 Finally, we note that both the redshift and the stellar-mass distributions of the 28,993 \hii\ subcubes (shown in Figure~\ref{fig:distall}), are similar to those of the 11,419 galaxies.

\subsection{The star-forming main-sequence}
\label{ssec:detection:mainseq}

An important question is whether our sample is dominated by  main-sequence galaxies, or whether it contains a significant population of starburst systems. To test this, we binned the sample of galaxies into multiple stellar-mass subsamples, and measured the average SFR in each of the stellar-mass bins to test whether the average SFR and average stellar mass are consistent with the star-forming main sequence at these redshifts. Further, the star-forming main sequence has been shown to evolve within our redshift range, $z=0.74-1.45$ \citep[e.g.][]{Whitaker14,Leslie20}. We hence divided our sample into two redshift ranges, $z = 0.74-1.00$ and $z= 1.00-1.45$, and, in each redshift bin, we measured the average SFR of galaxies in stellar-mass subsamples of width 0.5~dex. This was done by separately stacking the rest-frame 1.4~GHz luminosities of the galaxies in each stellar-mass subsample to determine their average SFR, using the procedure described in Section~\ref{sec:stackingCont}. The stacking was performed using weights such that the redshift distributions of the stacked sub-images in the stellar-mass subsamples are identical. Finally, for each redshift bin, we compared our measurements of the average stellar mass and the average SFR with the star-forming main-sequence relation of \citet{Whitaker14}. We note there were fewer than 35~galaxies with $\Ms > 10^{11} \  \Msun$ in both redshift intervals; the average SFRs of these highest-stellar mass subsamples may thus be affected by small number statistics and are hence not used for this comparison.

\begin{figure}
    \centering
    \includegraphics[width=0.6\linewidth]{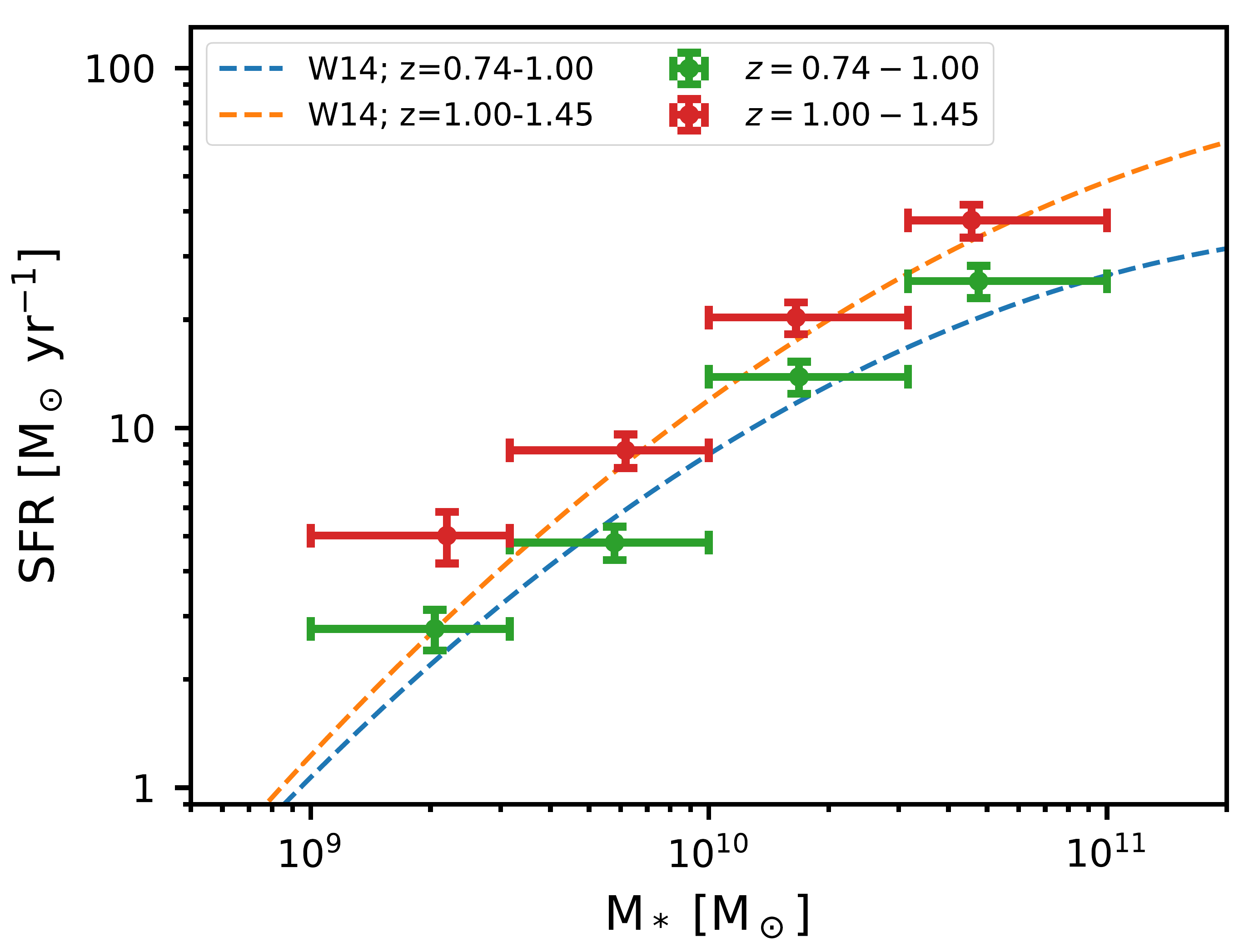}
    \caption{The average stellar-mass and average SFR of the DEEP2 galaxies of our sample. The filled circles show the average SFRs of galaxies in four stellar-mass subsamples at $z=0.74-1.00$ (green) and $z=1.00-1.45$ (red), obtained via stacking their rest-frame 1.4~GHz continuum emission. The vertical error bars show the $1\sigma$ error on each measurement, obtained by adding in quadrature the $1\sigma$ statistical uncertainty and the $10\%$ flux-scale uncertainty. The horizontal error bars show the range of stellar masses in each subsample. The dashed blue and orange curves (``W14'') show the star-forming main-sequence relation obtained for a stellar-mass-complete sample of star-forming galaxies in each redshift range \citep{Whitaker14}. The average SFRs and average stellar masses of our galaxies are seen to be consistent with the star-forming main-sequence relation in both redshift intervals.}
    \label{fig:det:mainsequence}
\end{figure}

Figure~\ref{fig:det:mainsequence} shows our measurements of the average SFR of galaxies in the four stellar-mass subsamples, for the two redshift bins, $z=0.74-1.00$ and $z=1.00-1.45$. For comparison, we also plot the star-forming main-sequence relations obtained for a stellar-mass-complete sample of star-forming galaxies at similar redshifts \citep{Whitaker14}. These authors provide the star-forming main-sequence relation for the redshift ranges $z=0.5-1.0$ and $z=1.0-1.5$; we have interpolated between these measurements to infer the main-sequence relations at $z=0.74-1.00$ and $z=1.00-1.45$. Figure~\ref{fig:det:mainsequence} shows that our average SFR values in each of the four stellar-mass bins are consistent with the star-forming main-sequence relations in the two redshift intervals.

\section{The average \hi\ mass and the optimum spatial resolution}
\label{sec:massresolution}

\begin{figure}
    \centering
    \includegraphics[width=\linewidth]{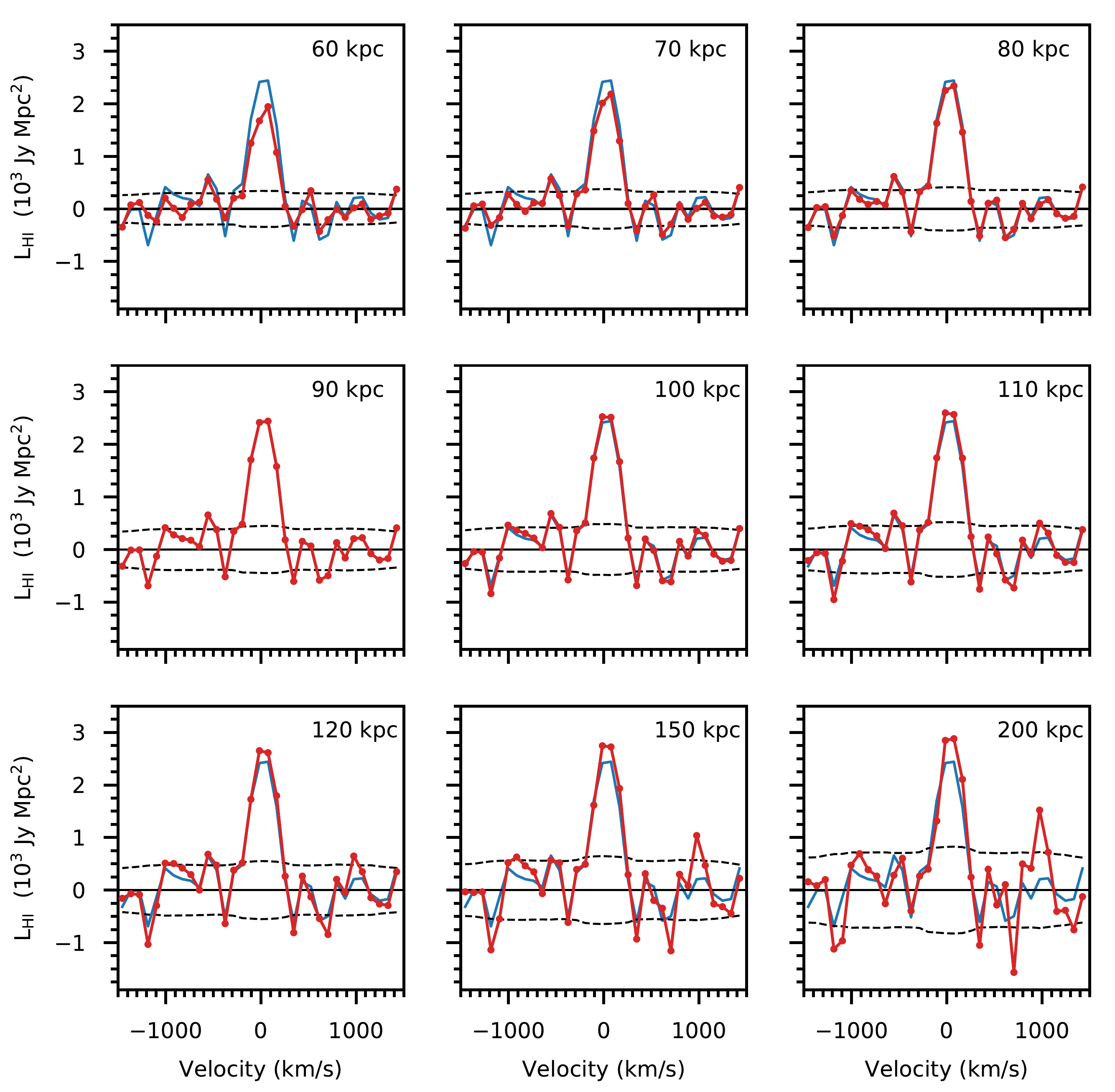}
    \caption{The stacked \hii\ emission spectra of the DEEP2 galaxies, obtained at different spatial resolutions, from 60~kpc to 200~kpc. Each panel shows (in red) the stacked \hii\ spectrum at a particular resolution, listed at the top right, with the black dashed curves indicating the $\pm1\sigma$ error on each 90~\kmps\ channel. For comparison, the stacked \hii\ spectrum at a resolution of 90~kpc, the optimal spatial resolution, is shown in blue in all panels.}
    \label{fig:HIspectrares}
\end{figure}

\begin{table}
    \centering
    \begin{tabular}{|c|c|c|c|}
    \hline \hline 
        Spatial Resolution  & $\langle \MHI \rangle$  & SNR \\
         (kpc) & ($\times 10^9~\Msun$) & \\
    \hline 
    60 & 10.0 $\pm$ 1.5 & 6.7 \\
    70 & 11.7 $\pm$ 1.6 & 7.3 \\
    80 & 12.9 $\pm$ 1.8 & 7.4\\
    90 & 13.7 $\pm$ 1.9 & 7.1\\
    100 & 14.2 $\pm$ 2.1 & 6.8 \\
    110 & 14.6 $\pm$ 2.3 & 6.4 \\
    120 & 14.8 $\pm$ 2.4 & 6.2 \\
    150 & 15.2 $\pm$ 2.8 & 5.4 \\
    200 & 15.4 $\pm$ 3.7 & 4.2 \\
 \hline \hline
    \end{tabular}
    \caption{The average \hi\ mass of the DEEP2 galaxies of our sample at different spatial resolutions, from 60~kpc to 200~kpc. The average \hii\ emission within the central beam was obtained from the stacked \hii\ spectra of Figure~\ref{fig:HIspectrares}. The last column lists the SNR of the detection of the stacked \hii\ emission signal.}
    \label{tab:HImassres}
\end{table}
\begin{figure}
    \centering
    \includegraphics[width=0.5\linewidth]{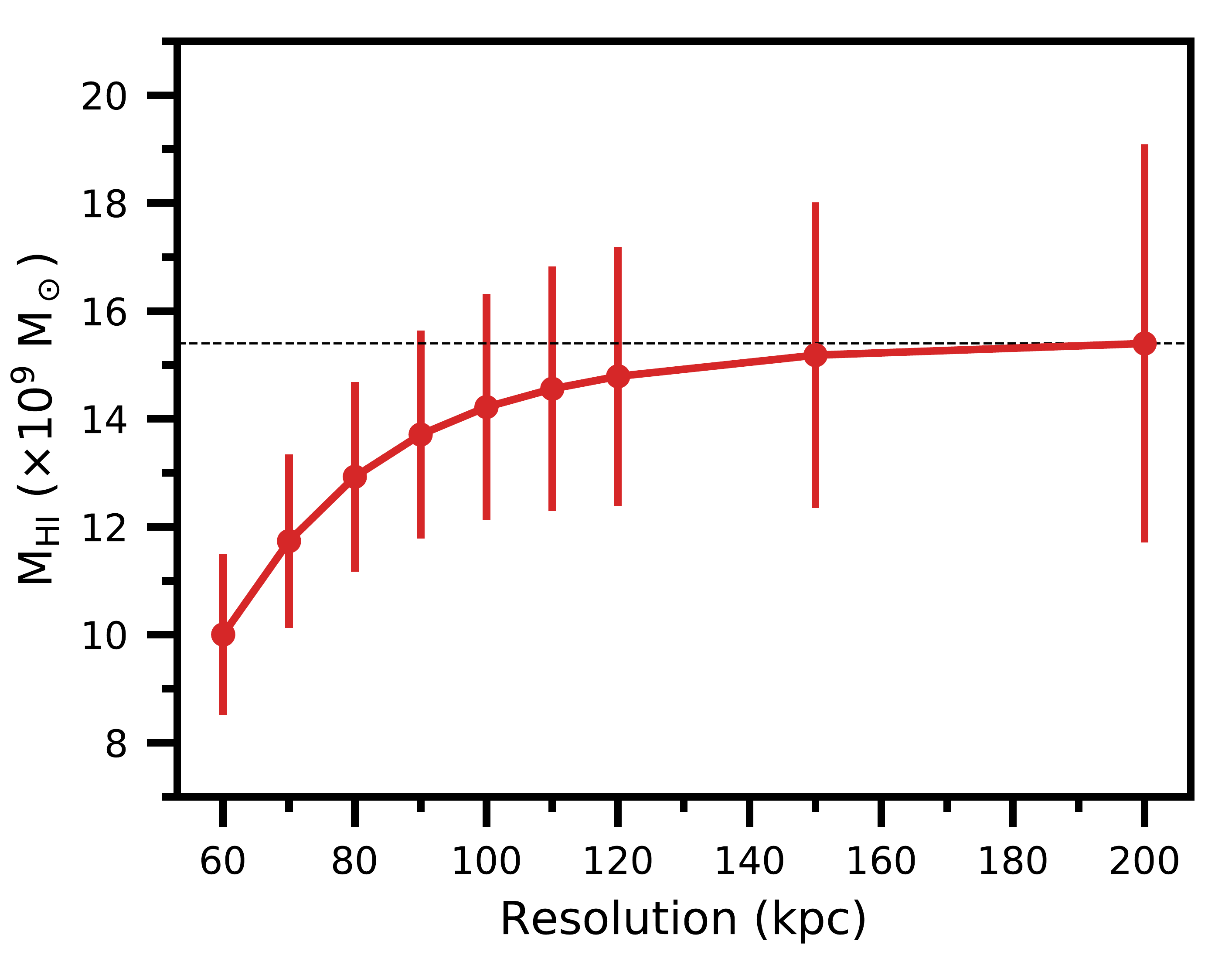}
    \caption{The average \hi\ mass of the DEEP2 galaxies of our sample, plotted as a function of spatial resolution. The dashed line shows the average \hi\ mass at a spatial resolution of 200~kpc.}
    \label{fig:HImassres}
\end{figure}

We used the procedure described in Section~\ref{sec:stackingHI} to stack the 28,993 \hii\ subcubes of the 11,419 blue star-forming galaxies of our sample, at the nine different spatial resolutions (60~kpc, 70~kpc, 80~kpc, 90~kpc, 100~kpc, 110~kpc, 120~kpc, 150~kpc, and 200~kpc) at which we obtained \hii\ subcubes of the DEEP2 galaxies. For each resolution, the \hii\ stacking was carried out with identical weights assigned to each \hii\ subcube. Figure~\ref{fig:HIspectrares} shows the stacked \hii\ spectra of the 11,419 galaxies at the nine different spatial resolutions. We clearly detect the average \hii\ emission signal at all spatial resolutions, with $\approx4.2\sigma-7.4\sigma$ significance.

We integrated the detected \hii\ emission signals of Figure~\ref{fig:HIspectrares} over the velocity range [$-150$~\kmps, +210~\kmps] to infer the average \hi\ mass of the DEEP2 galaxies, within the central spatial beam; these measurements are listed in Table~\ref{tab:HImassres}. Figure~\ref{fig:HIspectrares} shows the average \hi\ mass of the galaxies as a function of spatial resolution.  The average \hi\ mass is seen to increase from $(10.0 \pm 1.5)\times10^{9} \ \Msun$ at a resolution of 60~kpc to $(15.4 \pm 3.7)\times10^{9} \ \Msun$ at 200~kpc. Further, the increase in the average \hi\ mass shows clear evidence of flattening at coarser resolutions (see Fig.~\ref{fig:HImassres}). Increasing the size of the spatial beam of the \hii\ subcubes comes at the cost of down-weighting the longer GMRT baselines, and consequently raising the RMS noise on the stacked spectrum. We find that the RMS error on the stacked \hii\ cube increases approximately linearly with spatial resolution over $60-200$~kpc, with the SNR of the detections peaking at a resolution of $\approx80$~kpc (see Table~\ref{tab:HImassres}).

\subsection{The Optimal Spatial Resolution}
The choice of spatial resolution is important in terms of both accurately measuring the average \hi\ mass of the DEEP2 galaxies and maximizing the SNR of the final \hii\ spectrum. As can be seen from Table~\ref{tab:HImassres} and Figure~\ref{fig:HImassres}, using too narrow a spatial resolution resolves out some of the \hii\ emission signal, and results in under-estimating the average \hi\ mass. Conversely, using too coarse a spatial resolution would include all the \hii\ emission but at the cost of increasing the RMS noise on the signal. Ideally, the spatial resolution would be matched to the spatial extent of the \hii\ emission. However, the spatial extent of the \hii\ emission from galaxies at $z\approx1$ is not {\it a priori} known. We hence aimed to identify the optimal spatial resolution by measuring the average \hi\ mass at different resolutions, and finding the resolution above which the measured average \hi\ mass does not continue to increase.


An important subtlety in measuring the average \hi\ mass at different spatial resolutions arises from the fact that the mass measurements are correlated, as a subset of the interferometer visibilities is common to the measurements. Thus, one cannot assume independent errors when taking the difference between the values of the average \hi\ mass measured at different spatial resolutions. We hence measured the difference between the average \hi\ masses at each pair of spatial resolutions by stacking the difference of the \hii\ spectra at the location of each galaxy at the two resolutions. For example, to estimate the excess \hi\ mass at 80-kpc resolution relative to 60-kpc resolution, we first subtracted the \hii\ spectrum of each individual subcube at 60-kpc resolution from that of the same subcube at 80-kpc resolution to obtain a difference spectrum for the subcube. We then stacked these difference spectra to measure the excess \hii\ emission signal. The error on the stacked difference spectrum for each pair of resolutions was obtained via the Monte Carlo approach described in Section~\ref{sec:stackingHI}. The above procedure was carried out for every pair of the spatial resolutions at which we obtained the \hii\ subcubes, i.e. every pair of 60~kpc, 70~kpc, 80~kpc, 90~kpc, 100~kpc, 110~kpc, 120~kpc, 150~kpc, and 200~kpc.


We find evidence at $\gtrsim 2.5\sigma$ statistical significance for an increase in the average \hi\ mass of our sample of 11,419 galaxies out to a resolution of 90~kpc. However, the difference between the average \hi\ mass measured at a spatial resolution of 90~kpc and at any coarser resolution has $< 2\sigma$ significance. In other words, the average \hi\ mass of our galaxies, measured at a resolution of $90$~kpc is consistent, within statistical uncertainties, with that measured at coarser resolutions.  We thus identify 90~kpc as the optimal spatial resolution for the \hii\ stacking.

Unless otherwise mentioned, all subsequent results of the \hii\ emission stacking of this survey are at a spatial resolution of 90~kpc.


\section{The Average \hi\ properties of Star-Forming Galaxies at $z\approx1$}
\label{sec:histack}

In this section, we present the main results of this study, from stacking the \hii\ line emission and the rest-frame 1.4~GHz continuum emission signals of the full sample of 11,419 blue star-forming galaxies. Possible systematic effects are considered in Section~\ref{sec:systematic}; we note here, in passing, that we find no evidence in Section~\ref{sec:systematic} for systematic effects that might affect our results.

\begin{table}
\centering

\begin{tabular}{|c|c|}
\hline
    \hline
    Number of Galaxies & $11,419$ \\
     \hline
    Number of \hii\ subcubes & $28,993$ \\
\hline
    Redshift range & $0.74-1.45$ \\
\hline
    Mean redshift, $\langle z \rangle$ & $1.04$ \\
\hline
    Stellar mass range & $1.0 \times 10^9 \ \Msun \ - 2.4 \times 10^{11} \ \Msun$ \\
    \hline
    Mean stellar mass, $\langle \Ms \rangle$ & $9.9 \times 10^9 \ \Msun$\\
\hline
    Mean Radio-derived SFR & $8.07 \pm 0.82 \ \Msun/\textrm{yr}$ \\
     \hline
     Mean \hi\ mass,  $\langle \MHI \rangle$  & $(13.7 \pm 1.9) \times  10^{9}~\Msun$ \\
     \hline
    
    \hline
    \hi\ depletion timescale, $\langle \tdephi \rangle$  & $(1.70 \pm 0.29)$~Gyr \\
     \hline
     \hline
\end{tabular}
\caption{ Details of the sample. The rows are (1)~the number of galaxies, (2)~the number of independent \hii\ subcubes, (3) the redshift range of the galaxies, (4)~their average redshift, $\langle z \rangle$, (5)~the stellar mass range of the galaxies, (5)~their average stellar mass, $\langle \Ms \rangle$, (6)~the SFR derived from the rest-frame average 1.4~GHz radio luminosity density, (7)~the average \hi\ mass, $\langle \MHI \rangle$, and (8)~the characteristic \hi\ depletion timescale, $\langle\tdephi\rangle\equiv\langle \MHI \rangle/\langle\textrm{SFR}\rangle$. See main text for discussion.  } 
\label{tab:avgprop}
\end{table}

\subsection{The average \hi\ Mass and the average SFR of the Sample}
\label{ssec:himass}

\begin{figure}
    \centering
    \includegraphics[width=\linewidth]{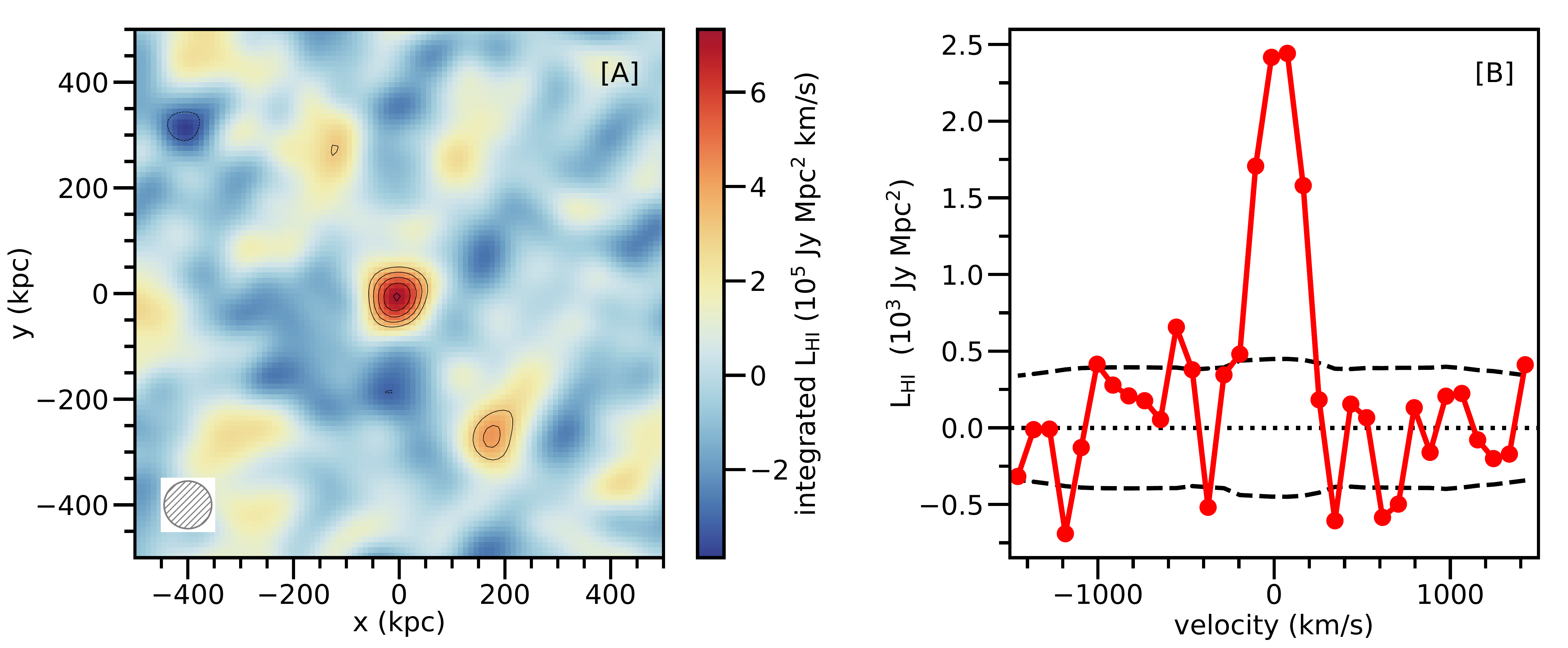}
    \caption{The average \hii\ emission signal from blue star-forming galaxies at $z\approx1$. [A]~The average \hii\ emission image of the 11,419 blue star-forming galaxies at $z=0.74-1.45$, obtained by integrating the emission over the central velocity channels. The circle on the bottom left of the panel shows the 90-kpc spatial resolution of the image. The contour levels are at $-3.0\sigma$ (dashed), $+3.0\sigma$, $+4.0\sigma$, $+5.0\sigma$, $+6.0\sigma$,  and $+7.0\sigma$ statistical significance. The average \hii\ emission signal is clearly detected in the central region of the image. The effect of the synthesized beam on the noise properties of this stacked image is discussed in Section~\ref{ssec:deconv} and Figure~\ref{fig:deconv}. [B]~The average \hii\ emission spectrum of the 11,419 blue star-forming galaxies at $z=0.74-1.45$, obtained by taking a spectral cut through the location of the peak luminosity density of the left panel. The channel width of the spectrum is 90~\kmps. The dashed horizontal curves show the $\pm1\sigma$ error on the spectrum. A clear detection of the average \hii\ emission signal can be seen in the central velocity channels of the spectrum. }
    \label{fig:90kpcdet}
\end{figure}

\begin{figure}
    \centering
    \includegraphics[width=\linewidth]{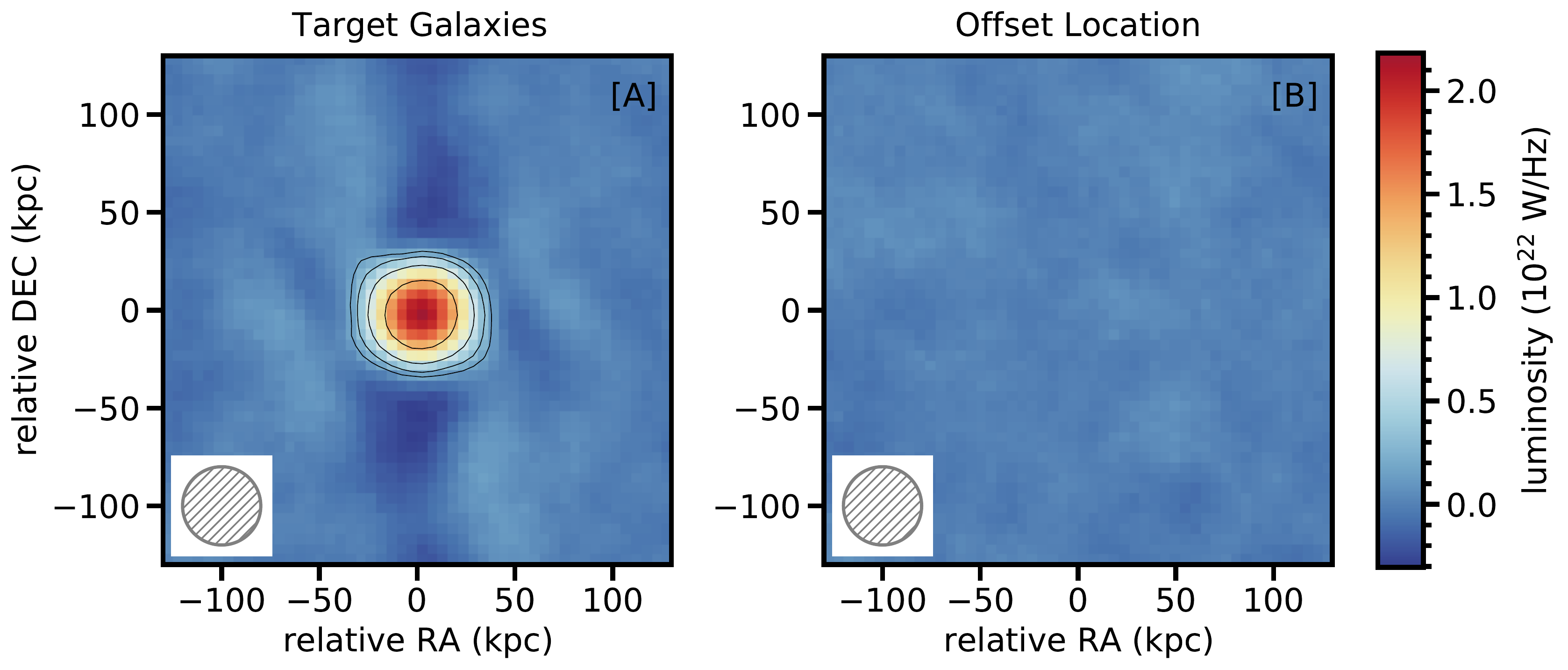}
    \caption{The median-stacked rest-frame 1.4~GHz continuum emission from [A]~the DEEP2 galaxies, and [B]~positions offset from the DEEP2 galaxies by $100\arcsec$.
    A clear ($\approx 67\sigma$ significance) detection is visible at the location of the stacked DEEP2 galaxies in the left panel, while the stack at offset positions in the right panel shows no evidence for either continuum emission or any systematic patterns. The contour levels are at $5\sigma,\ 10 \sigma,\ 20\sigma, \ {\rm and} \ 40\sigma$ statistical significance. The circle in the bottom left corner represents the $40$~kpc beam to which all continuum images were convolved before the stacking.}
    \label{fig:contstack}
\end{figure}

The stacked \hii\  emission image and the stacked \hii\ spectrum of the 11,419 blue star-forming galaxies, obtained by stacking the 28,993 independent \hii\ subcubes at a spatial resolution of 90~kpc, are shown in Fig.~\ref{fig:90kpcdet}. A clear detection of the stacked \hii\ emission signal can be seen in both the stacked spectrum and the stacked \hii\ emission image. The velocity-integrated average \hii\ signal has $7.1\sigma$ statistical significance. Integrating the stacked \hii\ spectrum over the velocity range [$-150$~\kmps, $+210$~\kmps], we find that the average velocity-integrated \hii\ line luminosity of the 11,419 galaxies at $\langle z \rangle = 1.04$ is $ (7.33 \pm 1.03) \times 10^{5}$~Jy~Mpc$^2$~km~s$^{-1}$. This yields an average \hi\ mass of $\langle\MHI \rangle = (1.37 \pm 0.19) \times  10^{10} \ \Msun$ for blue star-forming galaxies at $\langle z\rangle=1.04$. This is consistent with the earlier measurement of $\langle \MHI \rangle = (1.19 \pm 0.26) \times 10^{10} \ \Msun$ at $\langle z \rangle \approx 1.0$ by \citet{Chowdhury20}, from the 90-hours of observations in GMRT Cycle~35. 

The average SFR of the 11,419 galaxies of our sample was determined from their stacked rest-frame 1.4~GHz continuum luminosities, following the procedures of Section~\ref{sec:stackingCont}. Fig.~\ref{fig:contstack}[A] shows the image obtained by stacking the rest-frame 1.4~GHz luminosities of the 11,419 galaxies, while Fig.~\ref{fig:contstack}[B] shows the stack at positions offset from the DEEP2 galaxies: a clear detection of the stacked 1.4~GHz continuum emission, at $\approx 67\sigma$ significance, is visible in the left panel, while no systematic effects can be seen in the right panel. Using the SFR calibration of \citet{Yun01}, adjusted to a Chabrier IMF, the detection yields an average SFR of $(8.07\pm0.82)$~$\Msun \ \textrm{yr}^{-1}$ for the sample. The average properties of the blue, star-forming galaxies of our sample are summarised in Table~\ref{tab:avgprop}.

The \hi\ mass of blue galaxies in the local Universe, with NUV$-$r$<4$ and with a stellar mass distribution identical to the 11,419 galaxies at $z\approx1$, is $(3.96 \pm 0.17)\times10^{9}~\Msun$ \citep{Catinella18}\footnote{The errors on the average \hi\ mass of the xGASS galaxies from \citet{Catinella18} were computed using bootstrap resampling with replacement}. { We emphasize that the average \hi\ mass of the blue galaxies in the xGASS sample was computed with weights such that their effective stellar mass distribution is identical to that of the 11,419 blue star-forming galaxies at $z=0.74-1.45$.} As can be seen in Fig.~\ref{fig:redshiftevol}[A], we find clear evidence, at $5.1\sigma$ significance, for redshift evolution in the average \hi\ mass of blue galaxies beteween  $z\approx 1$ and $z\approx 0$. Specifically, we find that blue star-forming galaxies at $z \approx 1$, i.e. at the end of the epoch of peak cosmic SFR density, have \hi\ reservoirs that are larger by a factor of ($3.5\pm0.5$) than those of blue galaxies with an identical stellar-mass distribution in the local Universe.

\subsection{The \hi\ depletion timescale}
\label{sec:taudep}

\begin{figure}
    \centering
    \includegraphics[width=\linewidth]{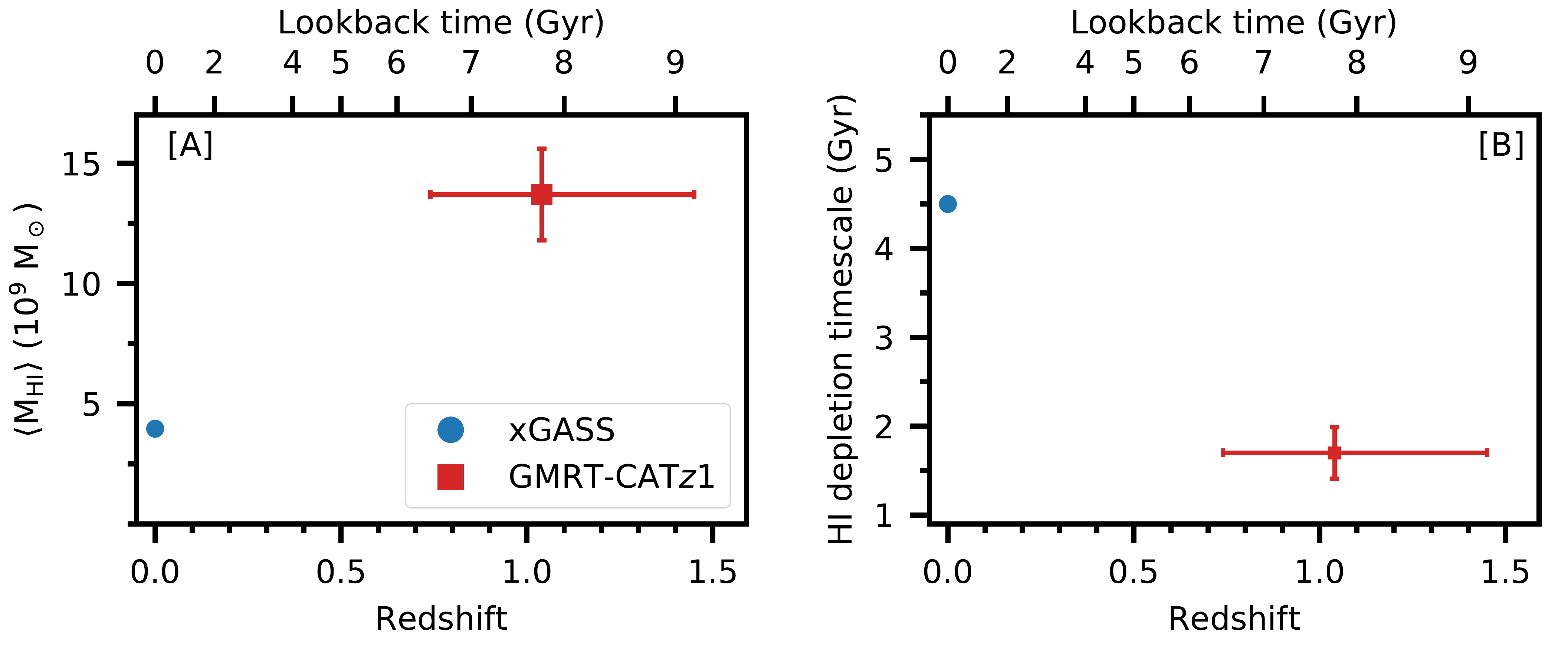}
    \caption{Redshift evolution of [A] the average \hi\ mass, $\langle\MHI\rangle$, and [B] the characteristic \hi\ depletion timescale, $\langle\MHI\rangle/\langle\textrm{SFR}\rangle$, for blue star-forming galaxies, having identical stellar mass distributions, with $\langle\Ms\rangle\approx10^{10}~\Msun$. In both panels, the blue circle shows the result from the xGASS survey of local Universe galaxies \citep{Catinella18}, while the red square shows the measurement at $z\approx1$ from the current work. }
    \label{fig:redshiftevol}
\end{figure}

The availability of neutral gas, the fuel for star-formation, determines the time for which a galaxy can sustain its current SFR. The gas depletion timescale, defined as the ratio of the gas mass (\hi\ or $\htwo$) to the SFR, is an important metric to understand how long a galaxy can sustain its current SFR. { We define the ``characteristic'' \hi\ and $\htwo$ depletion timescales of the galaxies of a given sample as, respectively, $\langle \tdephi \rangle \equiv\langle \MHI \rangle /\langle \textrm{SFR}\rangle$ and $\langle \tdephtwo \rangle \equiv\langle\MHtwo\rangle/\langle\textrm{SFR}\rangle$. In the local Universe, blue galaxies
with the same stellar mass distribution as the 11,419 galaxies in our sample have  $\langle \tdephi \rangle = (4.5 \pm 0.2)$~Gyr \citep{Catinella18} and $\langle \tdephtwo \rangle = (0.45\pm0.04)$~Gyr \citep{Saintonge17}\footnote{We have divided the molecular gas masses of \citet{Saintonge17} by a
	factor of 1.36 to obtain the $\htwo$ masses of their galaxies. Further, we note that the average gas depletion timescales in this paper are all calculated consistently by taking the ratio of the weighted-average gas mass to the weighted-average SFR of the sample, with weights such that the stellar mass distribution of the sample is identical to that of Figure~\ref{fig:distall}[B].}.}
The \hi\ depletion timescale of galaxies with $\Ms\approx10^{10}~\Msun$ in the local Universe is thus much longer than the $\htwo$\ depletion timescale. Local Universe galaxies can thus sustain their current SFR for a long timescale, $\langle \tdephi \rangle \approx 4.5$~Gyr,  even without the accretion of fresh gas from the CGM or via mergers. In other words, the availability of \hi\ is not a bottleneck in the star-formation process in local Universe galaxies.


For the DEEP2 galaxies, we combine the average \hi\ mass of $\langle \MHI \rangle = (13.7 \pm 1.9) \times 10^9 \ \Msun$ with the average SFR of $(8.07 \pm 0.82) \ \Msun \ \textrm{yr}^{-1}$ to obtain a characteristic \hi\ depletion timescale of $\langle \tdephi \rangle = (1.70 \pm 0.29)$~Gyr, for blue star-forming galaxies with $\langle \Ms \rangle \approx10^{10}~\Msun$ at $\langle z \rangle=1.04$. Our measurement of $\langle \tdephi \rangle$ of blue galaxies at $z\approx1$ is three times smaller than the characteristic \hi\ depletion timescale of $\langle \tdephi \rangle \approx 4.5 \pm 0.2$~Gyr of blue galaxies with an identical stellar mass distribution in the local Universe (see Figure~\ref{fig:redshiftevol}[B]). Further, the \hi\ depletion timescale at $z \approx 1$ is only a factor of $\approx 3$ higher than the $\htwo$ depletion timescale of $\approx0.5$~Gyr\footnote{We note that this is the $\htwo$ depletion timescale, after dividing the molecular gas depletion timescale by a factor of 1.36.} in main-sequence galaxies at similar redshifts \citep{Tacconi13}. Our results thus clearly establish, consistent with the findings of \citet{Chowdhury20}, that blue star-forming galaxies at the end of the epoch of peak cosmic SFR density can sustain their SFR for a short timescale,, only $\approx1.5$~Gyr, in the absence of gas accretion from the CGM. This supports the hypothesis that accretion of gas from the CGM may have been insufficient to sustain the high SFR of galaxies at $z\approx1$, causing the observed decline in the star-formation activity of the Universe at lower redshifts \citep{Bera18,Chowdhury20,Chowdhury21}. {Further, using \hii\ data from the present GMRT-CAT$z$1 survey, \citet{Chowdhury22} find direct evidence that the average \hi\ mass of star-forming galaxies at $z \approx 1$ is a factor of $\approx 3.2$ lower than that of galaxies with the same  stellar-mass distribution at $z \approx 1.3$, indicating that accretion was insufficient to replenish the gas reservoirs of massive galaxies.}

{In passing, we note that there are a number of possible causes for the decline in the gas accretion on to massive galaxies at $z \lesssim 1$, including (1)~a transition in the mode of accretion from cold mode at high redshifts to hot mode at $z \lesssim 1$, which would slow down the accretion process \citep[e.g.][]{Keres05,Dekel09a}, (2)~AGN or stellar feedback,  especially in massive galaxies \citep[e.g.][]{Weiner09,Steidel10,Kakkad20,Valentino21}, and (3)~heating of the gas reservoir in the CGM \citep[e.g.][]{Schawinski14}. The current \hii\ (and ancillary) data on the DEEP2 fields do not allow us to distinguish between these (and other) possibilities. Further, it is also possible that environmental effects \citep[e.g.][]{Tal14} might contribute to the decline in the star-formation activity at $z \lesssim 1$. Deeper \hii\ data should allow us to separate the galaxies into isolated systems and groups, and enable us to study such environmental effects.}


{ \section{The Average \hi\ Mass of Red Galaxies and AGNs at $z\approx1$}
\label{sec:redagn}
\subsection{Red Galaxies}
\begin{figure}
    \centering
    \includegraphics[width=\linewidth]{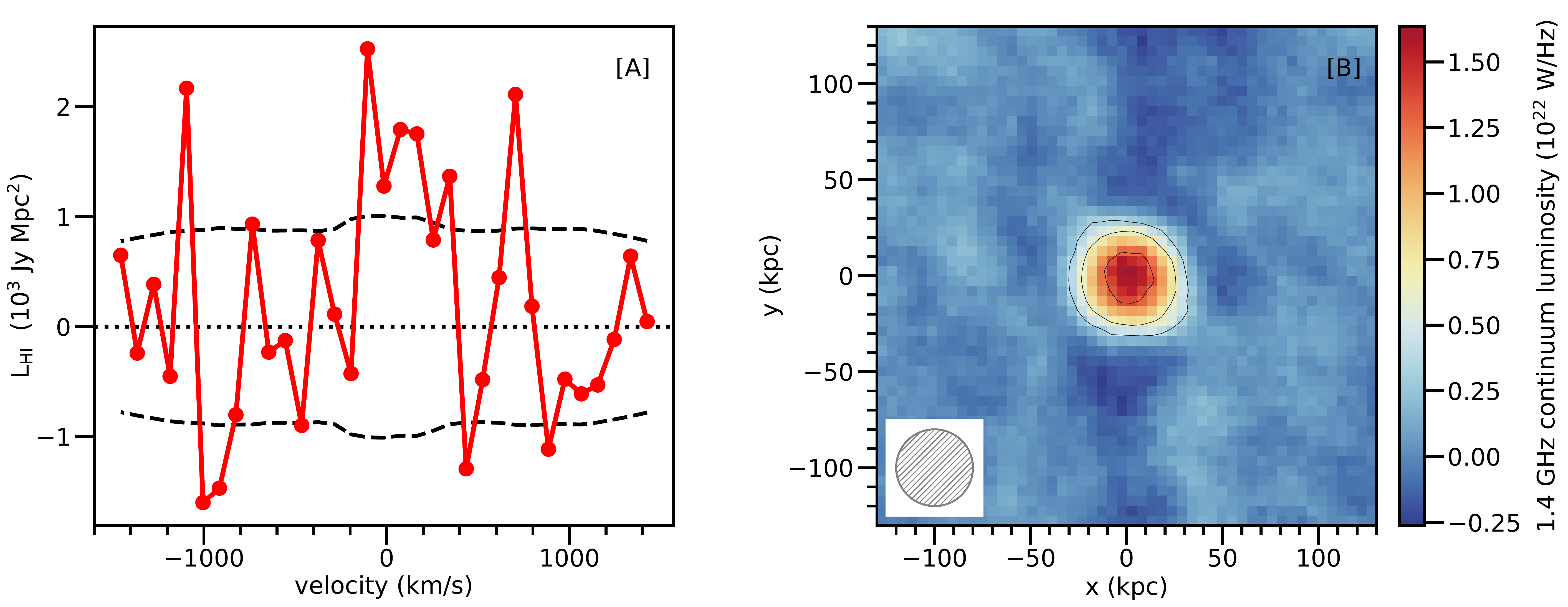}
    \caption{[A] The stacked \hii\ spectrum from the 1,738 red galaxies at $z=0.74-1.45$, obtained by stacking the \hii\ spectra obtained at the location of the galaxies. The channel width of the spectrum is 90~\kmps. The dashed horizontal curves show the $\pm1\sigma$ error on the spectrum. A tentative  ($2.9\sigma$ significance) detection of the average \hii\ emission signal can be seen in the central velocity channels of the spectrum. [B]~The median-stacked rest-frame 1.4~GHz continuum emission from the red galaxies of the left panel. A clear ($\approx 25\sigma$ significance) detection of the stacked continuum emission is visible at the centre of the image, i.e. at the location of the stacked DEEP2 galaxies. The contour levels are at $5\sigma,\ 10 \sigma,\ {\rm and} \ 20\sigma$ statistical significance. The circle in the bottom left corner represents the $40$~kpc beam to which all continuum images were convolved before the stacking.}
    \label{fig:redstacks}
\end{figure}
The DEEP2 survey targetted galaxies down to a limiting magnitude of R$_\textrm{AB}=24.1$; this results in a bias against red galaxies at $z>0.7$, with the bias becoming stronger at higher redshifts (see Section~\ref{sec:sample}; \citealp{Willmer06,Newman13}). Indeed, only $\approx14\%$ of the DEEP2 galaxies at $z=0.74-1.45$ that lie within our uGMRT pointings are part of the ``red cloud'' (see Section~\ref{sec:sample}) and only $\approx 26\%$ of these are at $z>1$. In order to ensure the homogeneity of our sample, we had excluded the red DEEP2 galaxies from our main sample of blue star-forming galaxies. Here, we examine the average \hi\ mass and the average SFR of the red DEEP2 galaxies at $z=0.74-1.45$, lying within our uGMRT pointings, and compare their average properties to those of the blue DEEP2 galaxies of our main sample.

Our GMRT observations cover the redshifted \hii\ line for 2,222 red galaxies at $z=0.74-1.45$. After excluding galaxies hosting radio-bright AGNs (see Section~\ref{ssec:samplesel}) and galaxies whose \hii\ subcubes were affected by systematic effects (identified using the procedures of Section~\ref{ssec:spectraflag}), our sample contains 4,346 independent \hii\ subcubes of 1,738 red galaxies at $z=0.74-1.45$. 

We stacked the 4,346 \hii\ subcubes of the 1,738 red galaxies at $z=0.74-1.45$, following the procedures of Section~\ref{sec:stackingHI}. Figure~\ref{fig:redstacks}[A] shows the stacked \hii\ emission signal of the 1,738 red galaxies. We obtain a tentative detection, at $2.9\sigma$ statistical significance, of the average \hii\ emission from the 1,738 red galaxies at $\langle z\rangle=0.94$. Integrating the average \hii\ emission signal  over the velocity range [-150~\kmps,+210~\kmps] yields an average \hi\ mass of $\langle\MHI\rangle=(12.4\pm4.2)\times10^9~\Msun$. The average \hi\ mass of the red galaxies is thus consistent with $\langle\MHI \rangle = (13.7 \pm 1.9) \times  10^{9} \ \Msun$, the average \hi\ mass of the 11,419 blue star-forming galaxies of our main sample (Section~\ref{sec:histack}). However, we note that the average stellar mass of the 1,738 red galaxies is $\langle\Ms\rangle\approx6\times10^{10}~\Msun$, far higher than the average stellar mass of the blue galaxies, $\langle\Ms\rangle\approx10^{10}~\Msun$. The ratio of the average \hi\ mass to the average stellar mass, $\langle\MHI \rangle/\langle\Ms\rangle=0.21\pm0.07$, for the red galaxies is thus far lower than the same ratio, $\langle\MHI \rangle/\langle\Ms\rangle=1.38\pm0.19$, for the blue star-forming galaxies. 

However, we note that both the stellar-mass and the redshift distributions of the samples of blue galaxies and red galaxies are different, with the red sample having both higher stellar masses and lower redshifts. {\citet{Chowdhury22} find that the average \hi\ properties of blue galaxies depends on both their average redshift and their average stellar mass; these dependences would affect the above comparisons between the red and blue galaxies. The differences in the redshift and the stellar-mass distributions of the two galaxy samples should be taken into account by using appropriate weights in future comparisons.}

We estimate the average SFR of the 1,738 red galaxies by median-stacking their rest-frame 1.4~GHz continuum luminosities, following the procedures of Section~\ref{sec:stackingCont}. Figure~\ref{fig:redstacks}[B] shows the stacked rest-frame 1.4~GHz continuum emission of the red galaxies. We clearly detect the stacked 1.4~GHz continuum emission from the sample of red galaxies, at $\approx25\sigma$~significance, obtaining an average SFR of $6.05\pm0.65~\Msun\textrm{yr}^{-1}$. As noted above, the sample of red galaxies has a high average stellar mass, $\langle\Ms\rangle\approx6\times10^{10}~\Msun$; the average SFR of $6.05\pm0.65~\Msun\textrm{yr}^{-1}$ for these galaxies is $\approx4.5$ times lower than that of main-sequence galaxies with $\langle\Ms\rangle\approx6\times10^{10}~\Msun$ at similar redshifts \citep{Whitaker14}. 
Combining the average \hi\ mass and the average SFR yields a characteristic \hi\ depletion timescale of $\langle \tdephi \rangle = (2.03 \pm 0.73)$~Gyr for the red galaxies at $\langle z \rangle = 0.94$.

 Overall, we find that the red galaxies in the DEEP2 survey are massive, but have an average SFR far lower than that of blue galaxies with similar stellar masses; {this suggests that most of the red objects are not dusty star-forming galaxies. The red galaxies also have a lower ratio of the average \hi\ mass to the average stellar mass than the 11,419 blue galaxies of our main sample, but also have a significantly higher stellar mass than the above blue galaxies.}


\subsection{Galaxies hosting AGNs}

The GMRT-CAT$z1$ survey covers the redshifted \hii\ line for 882 blue DEEP2 galaxies that were found to host an AGN with ${\rm L}_{\rm 1.4 GHz}>2\times10^{23}$~W/Hz (see Section~\ref{ssec:samplesel}). We investigated the 2,368 \hii\ subcubes of these 882 AGN-hosting galaxies for systematic issues, following the procedures described in Section~\ref{ssec:spectraflag}. After excluding the \hii\ subcubes affected by discernible systematic effects, we stacked the 2,087 \hii\ subcubes of the remaining 823 AGN-hosting galaxies. We do not detect the average \hii\ emission signal from this sample, obtaining a $3\sigma$ upper limit of  $\langle\MHI\rangle<24\times10^9~\Msun$ on the average \hi\ mass of the AGN-hosting galaxies, assuming an \hii\ line FWHM of $360$~\kmps. The upper limit on the \hi\ mass of the AGN-hosting galaxies is consistent with the average \hi\ mass of $\langle\MHI \rangle = (13.7 \pm 1.9) \times  10^{9} \ \Msun$ for our main sample of 11,419 blue star-forming galaxies with no AGNs. 



Overall, we find no evidence for a high average \hi\ mass in the galaxies of our sample that host AGNs. However, the number of such objects is more than an order of magnitude lower than the number of blue galaxies in our main sample, implying that we do not obtain tight constraints on the average \hi\ mass of the AGN sample. Probing the influence of AGNs on the average \hi\ properties of galaxies at $z \approx 1$ will require a larger AGN sample or significantly deeper observations.

}
\section{Tests for Systematic Issues in the \hii\ stacking}
\label{sec:systematic}

\subsection{The Effect of the Dirty Beam}
\label{ssec:deconv}
\begin{figure}
    \centering
    \includegraphics[width=\linewidth]{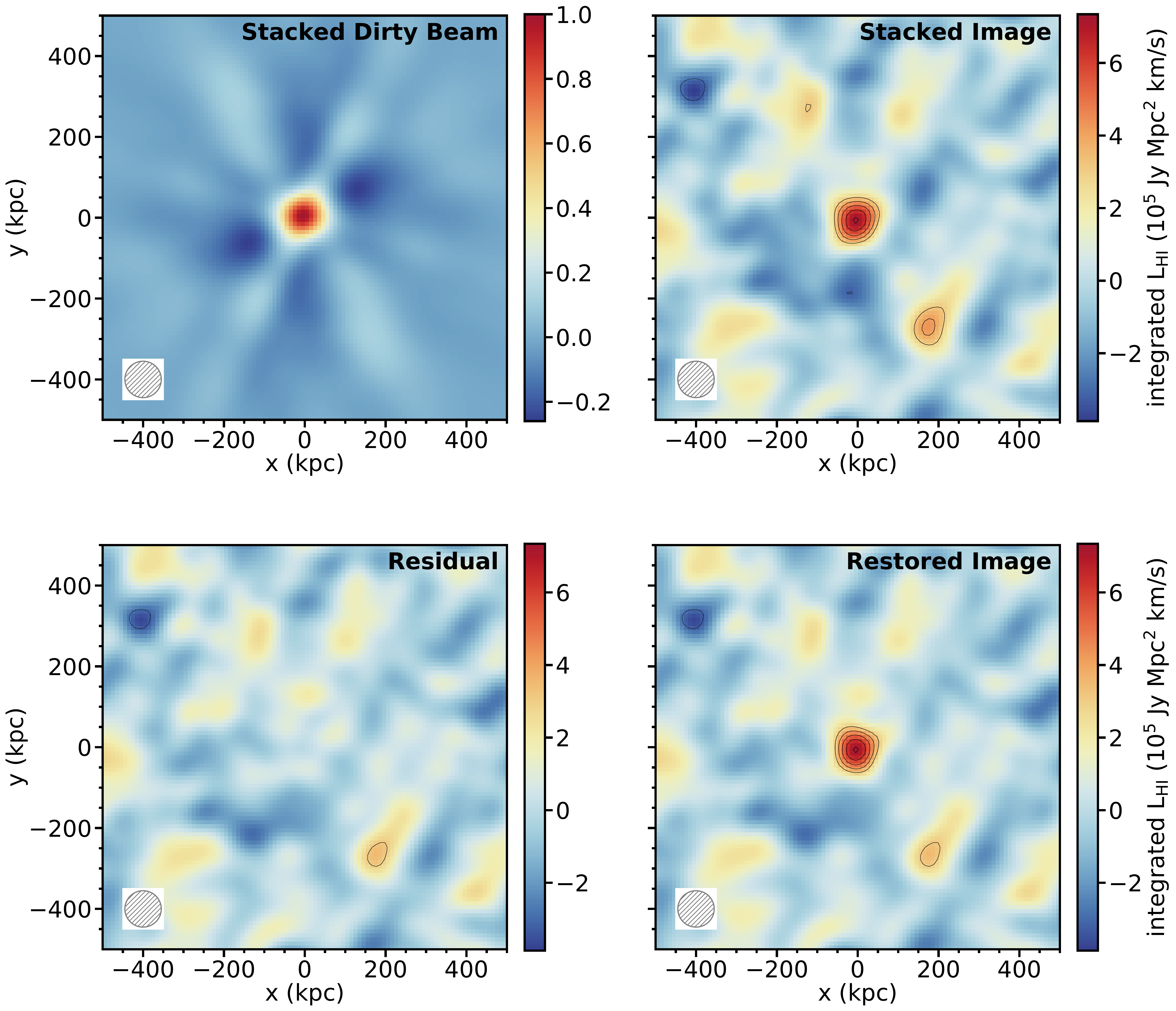}
    \caption{Subtraction of the dirty beam from the stacked \hii\ emission image of the 11,419 blue star-forming galaxies at $z=0.74-1.45$. The four panels show (clockwise, from top left): the stacked dirty beam, the stacked ``raw'' \hii\ emission image, the restored \hii\ emission image, and the residual image after subtracting out the scaled stacked dirty beam. The residual image was obtained by (a)~scaling the stacked dirty beam by the peak luminosity density of the stacked image, and  (b)~subtracting out the scaled dirty beam from the stacked image. The contour levels on all stacked \hii\ images are at $-3.0\sigma$ (dashed), $+3.0\sigma$, $+4.0\sigma$, $+5.0\sigma$, $+6.0\sigma$,  and $+7.0\sigma$ statistical significance. }
    \label{fig:deconv}
\end{figure}

The stacked \hii\ emission image of Fig.~\ref{fig:90kpcdet}[A] is the average of the observed \hii\ emission images of the individual 11,419 galaxies. Further, the observed \hii\ emission image of each galaxy is a convolution of the ``true'' \hii\ emission image with the point spread function (the ``dirty beam'') of the GMRT observations of the galaxy. 
The stacked \hii\ emission image of Fig.~\ref{fig:90kpcdet}[A] thus contains the combined effect of the point spread functions of the different observations. For normal \hi\ images (i.e. without stacking), in the absence of deconvolution, the point spread function would affect the noise properties of the image (yielding structures similar to those of the dirty beam). The fact that we are stacking \hii\ emission from different regions of the sky, with different dirty beams, implies that it is not straightforward to deconvolve the dirty beam from the stacked \hii\ emission image. However, the pattern of the average dirty beam is clearly visible on inspecting the stacked \hii\ emission image. To correct for this, we subtracted out the stacked dirty beam from the stacked \hii\ emission image; a similar deconvolution strategy was also used by \citet{Chen21} in an  \hii\ emission stacking experiment targetting galaxies at $z\approx0$. We emphasise that the deconvolution of the dirty beam carried out here is an approximation, which is exact only for the case of identical \hi\ masses of all the galaxies of the sample.

We obtain the stacked dirty beam by taking the average of the dirty beams of the independent \hii\ subcubes in the sample. Next, we scale the average dirty beam to the peak luminosity density of the stacked \hii\ emission image, and subtract out this scaled dirty beam from the stacked emission image to obtain the residual image. We then obtain the final restored \hii\ emission image by adding to the residual image a symmetric Gausssian having an FWHM of 90~kpc and the peak luminosity density of the original \hii\ emission image. Figure~\ref{fig:deconv} shows the stacked dirty beam, the original \hii\ emission image (the ``dirty image''), the residual \hii\ image, and the restored, stacked \hii\ emission image. A comparison between the original and restored \hii emission images in Figure~\ref{fig:deconv} shows clearly that the subtraction of the stacked dirty beam from the stacked \hii\ image improves the noise properties of the image. Further, the residual map in Figure~\ref{fig:deconv} shows no evidence for any extended emission at the location of the DEEP2 galaxies, consistent with the findings of Section~\ref{sec:massresolution} that the average \hii\ emission signal is unresolved at a spatial resolution of 90~kpc.

\subsection{The RMS noise of the stacked spectra}
\begin{figure}
    \centering
    \includegraphics[width=0.5\linewidth]{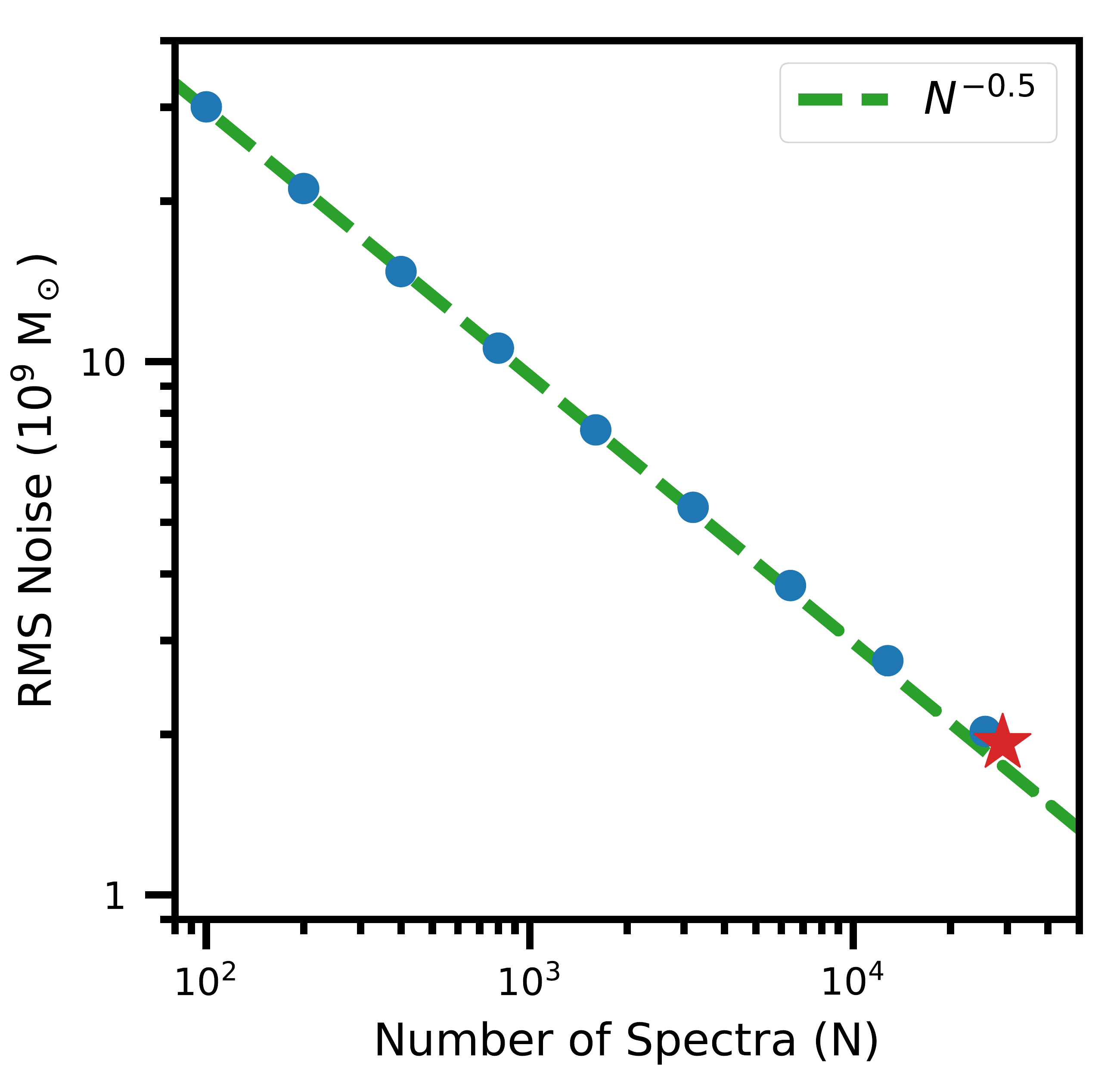}
    \caption{The RMS noise on the stacked \hii\ spectra as a function of the number of the \hii\ spectra that were stacked together. The blue points show the RMS noise, in units of \hi\ mass (assuming a velocity width of 360~\kmps), obtained on stacking N \hii\ spectra, randomly selected from the full sample. The red star shows the RMS noise on the stacked \hii\ spectrum using all the 28,993 \hii\ spectra in our sample. The dotted line shows the relation $\textrm{RMS} \propto \textrm{N}^{-0.5}$, expected if the noise properties of the individual \hii\ spectra are uncorrelated, with no underlying systematic effects. It is clear that the RMS noise obtained on the stacked spectra is consistent with the relation ${\rm RMS} \propto \textrm{N}^{-0.5}$. We thus find no evidence for systematic issues affecting the 28,993 \hii\ spectra of our sample.}
    \label{fig:det:rmsvar}
\end{figure}

The sensitivity of the stacking procedure critically depends on the noise properties of the individual \hii\ spectra that are stacked together. The spectral RMS noise on the stacked \hii\ spectrum is expected to decrease with the number (N) of individual \hii\ spectra as  RMS~$\propto \textrm{N}^{-0.5}$, for independent noise on the individual spectra. However, systematic issues in the data could introduce correlations between the individual spectra that may limit the sensitivity of the stacked spectrum. We tested for such issues in our sample of 28,993 \hii\ spectra by measuring the dependence of the RMS noise on the stacked spectrum on the number of spectra that were stacked together.

This was done by randomly selecting N spectra from our full sample of 28,993 \hii\ spectra, where N$=$100, 200, 400, 800, 1600, 3,200, 6,400, 12,800, and 25,600. For each N, we stacked the randomly-selected \hii\ spectra to obtain a stacked spectrum, and then followed the Monte~Carlo approach described earlier to estimate the RMS noise on the stacked spectrum. 

Figure~\ref{fig:det:rmsvar} shows the RMS noise obtained on the stacked \hii\ spectrum as a function of the number of stacked galaxies, N. It is clear from the figure that the RMS noise measurements are consistent with the relation RMS~$\propto \textrm{N}^{-0.5}$. We thus find no evidence for systematic issues that might limit the sensitivity of our stacked \hii\ spectrum.

 \subsection{The Effect of Source Confusion}
 
The stacked \hii\ emission signal can include,  in addition to the \hii\ emission from our target galaxies,  \hii\ emission from companion galaxies lying within the interferometer beam with \hii\ emission at the same velocities as the target galaxy. \citet{Chowdhury20} used the S$^{3}$-SAX-Sky \citep{Obreschkow09} simulations to find that the effect of such ``source confusion", due to companion galaxies lying within their 60~kpc spatial resolution, on the stacked \hii\ emission of galaxies at $z\approx1$ is small. We repeat the procedure of \citet{Chowdhury20}, but with a spatial resolution of 90~kpc, to find that the effect of source confusion on our measurement of the average \hi\ mass is also expected to be small, with companion galaxies contributing $\lesssim 5 \%$ to the observed average \hii\ emission signal.

We further probe the effect of source confusion on the measurement of the average \hi\ mass of our sample galaxies, by identifying target galaxies that have spectroscopically-identified companion galaxies in the DEEP2 DR4 catalogue \citep{Newman13}, such that the companion galaxies might contribute to our measurement of the \hii\ emission from the target galaxy. Our measurement of the average \hii\ emission is at a spatial resolution of 90~kpc, with the detected average \hii\ emission spanning $\approx\pm180$~\kmps\ around the systemic velocity. We find that only 276 galaxies of the 11,419 galaxies in our main sample ($\approx3\%$ of the sample) have spectroscopic companions in the DEEP2 DR4 catalogue that lie within $\pm45$~kpc and whose redshifts are within $\pm200$~\kmps\ of the target galaxy;  the \hii\ emission from these galaxies might be included in our measurement of the average \hii\ emission of the target galaxies. We exclude these 276 galaxies and stack the \hii\ subcubes of the  remaining 11,143 galaxies. This yields an average \hi\ mass for the 11,143 galaxies of $\langle\MHI \rangle = (1.40 \pm 0.20)  \times  10^{10} \ \Msun$. This is entirely consistent with the measured \hi\ mass of the full sample, $\langle\MHI \rangle = (1.37 \pm 0.19) \times  10^{10} \ \Msun$. We thus find no evidence that source confusion due to companion galaxies, identified in the DEEP2 spectroscopic catalogue, might affect our measurement of the average \hi\ mass of galaxies at $z\approx1$.

The spectroscopic completeness of the DEEP2 survey is $\approx50\%$ \citep{Conroy07}. The analysis presented above would thus not account for any companion galaxies for which the DEEP2 survey failed to obtain a redshift measurement. We estimate here the effect of such companion galaxies, not included in the DEEP2 spectroscopic catalogue, on our measurement of the average \hi\ mass. We find that 1,928 of the 11,419 galaxies in our sample have at least one object in the DEEP2 photometric catalogue \citep{Coil04} that lies within $\pm45$~kpc of the target galaxy and that meets the DEEP2 colour and magnitude selection criteria for spectroscopic targets in Fields 2, 3, and 4 \citep{Newman13}. Note that the redshift of the ``companion'' galaxies thus identified could be completely different from the redshift of the target galaxy. We exclude these 1,928 galaxies and stack the \hii\ subcubes of the remaining 9,491 galaxies of our sample to obtain an average \hi\ mass of $\langle\MHI \rangle = (1.48 \pm 0.21) \times  10^{10} \ \Msun$. This is again consistent with the measured \hi\ mass of the full sample, $\langle\MHI \rangle = (1.37 \pm 0.19) \times  10^{10} \ \Msun$. We thus find no evidence that source confusion due to companion galaxies, even those for which the DEEP2 survey failed to obtain a spectroscopic redshift, might affect our measurement of the average \hi\ mass of star-forming galaxies at $z\approx1$.

 Overall, we conclude that our measurement of the average \hi\ mass of galaxies at $z=0.74-1.45$ is not significantly affected by source confusion.

\section{Upper limits on the \hi\ Mass of Individual Galaxies}
\label{sec:uplims}
\begin{figure}
    \centering
    \includegraphics[width=\linewidth]{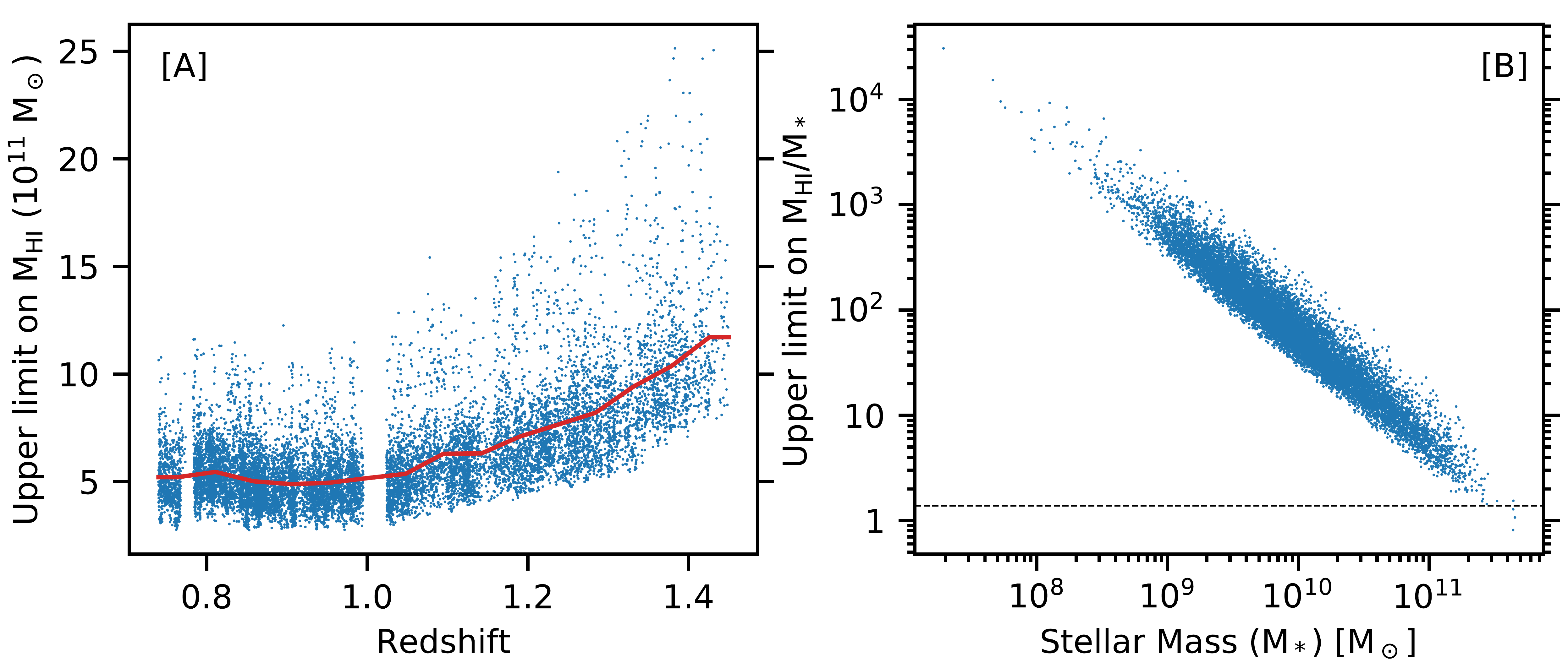}
    \caption{The 5$\sigma$ upper limits obtained on [A] the \hi\ mass and [B] the \hi\ fraction ($\MHI/\Ms$) of each of the 13,596 DEEP2 galaxies covered by the GMRT-CAT$z1$ survey. The upper limits, shown as blue points, assume an \hii\ line FWHM of 300~\kmps. The red curve in panel~[A] shows, as a function of redshift, the median of the $5\sigma$ upper limits on the \hi\ mass of the individual galaxies. The dotted horizontal line in panel~[B] shows the ratio of the average \hi\ mass to the average stellar mass for our main sample of 11,419 blue star-forming galaxies.  }
    \label{fig:masslim}
\end{figure}

We also carried out a search for \hii\ emission from the individual DEEP2 galaxies covered by the GMRT-CAT$z1$ survey at their spatial locations and redshifted \hii\ emission line frequencies. The search was carried out on all DEEP2 galaxies covered in our GMRT survey, including the 2,222 red galaxies, the  882 galaxies hosting radio AGNs, and the 487 galaxies with $\Ms<10^9~\Msun$  (see Section \ref{ssec:samplesel}). Further, we also retained the \hii\ subcubes of galaxies that were removed from our main sample due to the presence of a $\gtrsim 6\sigma$ feature in one of their subcubes (see Section~\ref{ssec:spectraflag}); this ensures that we do not exclude any real \hii\ emission signals. Overall, we searched for the \hii\ emission signal from 13,596 DEEP2 galaxies at $z=0.74-1.45$. 

For each DEEP2 galaxy, we first combined all available \hii\ subcubes, weighting by the inverse-variance of each cube, to obtain a single \hii\ subcube for the galaxy. The \hii\ subcubes used for the search had a spatial resolution of 90~kpc, the optimal resolution for the average \hii\ emission signal (see Section~\ref{sec:massresolution}). The search for \hii\ emission was carried out after smoothing the subcubes, using a boxcar kernel, to ten different velocity resolutions ranging from $30$~\kmps\ to $300$~\kmps, in steps of $30$~\kmps. The search was done at a range of velocity resolutions in order to maximize the sensitivity to \hii\ emission from galaxies at a range of inclinations. We searched for \hii\ emission, at $\geq5\sigma$ statistical significance, at the location of each DEEP2 galaxy and in the central $\pm250$~\kmps\ around its redshifted \hii\ emission frequency. We did not detect any emission feature with  $\ge5\sigma$ significance in the spectra of any of the 13,596 DEEP2 galaxies, at any of the ten velocity resolutions. 

The $5\sigma$ upper limits on the \hi\ masses of the 13,596 galaxies, for an assumed ``typical'' \hii\ line FWHM of $300$~\kmps, are shown in Figure~\ref{fig:masslim}[A]. We obtain a median $5\sigma$ upper limit of $\approx5\times10^{11} \ \Msun$ for galaxies at $z\approx0.74-1.0$, and $\approx(5-12)\times10^{11}~\Msun$ for galaxies at $z\approx1.0-1.45$. The corresponding upper limits on the \hi\ fraction ($\MHI/\Ms$) for each of the 13,596 galaxies are shown in Figure~\ref{fig:masslim}[B]. Except a few very massive galaxies, the upper limits on the \hi\ fraction of the 13,596 galaxies are far higher than our measurement of the ratio of the average \hi\  mass to the average stellar mass, $\langle\MHI\rangle/\langle\Ms\rangle=1.38\pm0.19$, of the sample. Assuming that the galaxies are viewed close to face-on, with an \hii\ line FWHM of $60$~\kmps, yields a median $5\sigma$ upper-limit of $\approx2.5\times10^{11}~\Msun$ for galaxies at $z\approx0.74-1.0$, and $\approx(2.5-5)\times10^{11}~\Msun$ for galaxies at $z\approx1.0-1.45$. These $5\sigma$ upper limits on the \hi\ mass of the 13,596 galaxies are $\approx10-100$ times higher than our measurement of the average \hi\ mass of blue galaxies at $z\approx1$ (see Section~\ref{sec:histack}). This allows us to rule out the presence of extremely large \hi\ reservoirs, with $\MHI\gtrsim(3-25)\times10^{11}~\Msun$, in any of the 13,596 DEEP2 galaxies at $z=0.74-1.45$.

\section{Summary}
\label{sec:conclusion}

In this paper, we present the GMRT-CAT$z1$ survey, a 510-hour GMRT \hii\ emission survey of galaxies at $z=0.74-1.45$ in seven sub-fields of the DEEP2 Galaxy Redshift Survey \citep{Newman13}. We describe the GMRT observations, the data analysis, and the main results obtained from stacking the \hii\ emission signals of our full sample of blue star-forming galaxies. Additional key results of the survey, including the role of \hi\ in the decline of star-formation activity of the Universe at $z \lesssim 1$, the contribution of \hi\ to the baryonic mass of galaxies at $z \approx 1$, the dependence of the \hi\ properties of star-forming galaxies at $z\approx1$ on their stellar properties, and estimates of the cosmological \hi\ mass density of the Universe at $z \approx 1$, will be described in separate papers \citep[e.g.][]{Chowdhury22,Chowdhury22b}. To summarise:

    


\begin{itemize}
    \item The GMRT observations cover the redshifted \hii\ line for 16,250 DEEP2 galaxies at $z=0.74-1.45$, lying within the half-power point of our GMRT pointings. We excluded red galaxies, radio AGNs, galaxies with $\Ms < 10^9 \ \Msun$, and any \hii\ subcubes affected by discernible systematic issues to obtain our main sample of 11,419 blue star-forming galaxies with $\Ms\ge10^{9}~\Msun$ at $z=0.74-1.45$. The observations provide up to 6 independent \hii\ subcubes spectra for each galaxy in the sample, yielding a total of 28,993 independent \hii\ subcubes for the 11,419 galaxies.
    
    \item To identify the optimal spatial resolution for the stacking, we stacked the \hii\ spectra of the 11,419 blue star-forming galaxies at nine different spatial resolutions, 60~kpc, 70~kpc, 80~kpc, 90~kpc, 100~kpc, 110~kpc, 120~kpc, 150~kpc and 200~kpc. We obtained clear detections of the average \hii\ signal at all nine resolutions, at $4.2-7.4\sigma$ statistical significance. We find that the average \hi\ mass of the sample at a resolution of 90~kpc is consistent with that at all coarser spatial resolutions. This implies that 90~kpc is the optimal spatial resolution for the \hii\ stacking for our galaxy sample. 
    
    \item We stacked the \hii\ subcubes of the sample at a resolution of 90~kpc to obtain a clear detection, with $\approx 7.1\sigma$ statistical significance, of the stacked \hii\ emission signal. The detection yields an average \hi\ mass of $\langle\MHI \rangle = (13.7 \pm 1.9) \times  10^{9} \ \Msun$ for blue star-forming galaxies with $\langle\Ms\rangle\approx10^{10}~\Msun$ at $\langle z \rangle = 1.04 $. 
    
    \item We stacked subsamples of galaxies to find that the RMS noise on the stacked \hii\ spectrum decreases with the number, N, of stacked spectra as $1/\sqrt{\rm N}$, as expected if the spectra have uncorrelated Gaussian noise. We thus find no evidence for any systematic effects that might affect our final stacked \hii\ spectrum.
    
    \item We investigated the effect of source confusion on our estimate of the average \hi\ mass of the sample, by excluding all  1,928 target galaxies with either  spectroscopic or photometric companions, with magnitudes and colours that meet the DEEP2 selection criteria, and that lie within $\pm 45$~kpc of a target galaxy. We stacked the \hii\ spectra of the remaining 9,491 galaxies, obtaining an average \hi\ mass consistent with that of the average \hi\ mass of the full sample of 11,419 galaxies. We thus find no evidence that the average \hi\ mass estimate might be contaminated by source confusion. We further used the S$^{3}$-SAX-Sky simulations \citep{Obreschkow09} to find that the effect of source confusion on our stacked \hii\ spectrum is expected to be $\lesssim 5\%$ at our spatial resolution of 90~kpc.
    
    \item We estimated the average SFR of the galaxies of the sample from their average rest-frame 1.4~GHz  luminosity, by carrying out a median stack of the rest-frame 1.4~GHz continuum emission. This yielded an average SFR of $(8.07\pm0.82)$~$\Msun\textrm{yr}^{-1}$. We also stacked the rest-frame 1.4~GHz continuum emission in galaxy subsamples based on stellar mass and redshift, to find that the average stellar masses and the average SFRs of the galaxies in each subsample are consistent with their lying on the main sequence at their respective redshifts.
    
    \item 
    We find that the average \hi\ mass of blue galaxies in the local Universe, with stellar-mass distribution identical to that of our full sample of 11,419 blue star-forming DEEP2 galaxies, is $(3.96\pm0.17)\times10^9~\Msun$. We thus find clear evidence, at $\approx 5.1\sigma$ significance, that the average \hi\ mass of blue galaxies has declined, by a factor of $(3.5\pm0.4)$, from $z\approx1$ to $z\approx0$.
    
    \item We combine our measurements of the average \hi\ mass and the average SFR of the 11,419 blue star-forming galaxies at $z\approx1$ to infer a characteristic \hi\ depletion timescale of $\langle\tdephi\rangle\equiv\langle \MHI \rangle/\langle\textrm{SFR}\rangle = (1.70 \pm 0.29)$~Gyr. This is significantly shorter than the characteristic \hi\ depletion timescale of $4.5 \pm 0.2$~Gyr in blue galaxies with an identical stellar-mass distribution in the local Universe. We thus confirm, at a  high statistical significance, the result that the \hi\ reservoir of star-forming galaxies at $z\approx1$ can sustain their current SFRs for a short period of $\approx1.7$~Gyr, in the absence of accretion of fresh atomic gas from the CGM or via minor mergers \citep{Bera18,Chowdhury20,Chowdhury21}. 
    
    \item  We stacked the \hii\ spectra of the 1,738 red galaxies at $z=0.74-1.45$ to obtain a tentative ($2.9\sigma$  significance) detection of their average \hii\ emission signal. We obtain an average \hi\ mass of $\langle\MHI\rangle=(12.4\pm4.2)\times10^9~\Msun$, consistent with the value of $\langle\MHI \rangle = (13.7 \pm 1.9) \times  10^{9} \ \Msun$ for the blue star-forming galaxies. However, the red galaxies have an average stellar mass of $6\times10^{10}~\Msun$, far higher than the average stellar mass of $10^{10}~\Msun$ of the blue galaxies. The ratio of the average \hi\ mass to average stellar mass of the red galaxies is $\langle\MHI \rangle/\langle\Ms\rangle=0.21\pm0.07$, far lower than the value $\langle\MHI \rangle/\langle\Ms\rangle=1.38\pm0.19$ of the blue galaxies. We also stacked the rest-frame 1.4~GHz luminosities of the 1,738 red galaxies to estimate an average SFR of $6.05\pm0.65~\Msun\textrm{yr}^{-1}$, $\approx4.5$ times lower than that of main-sequence galaxies at $z \approx 1$ with $\Ms\approx6\times10^{10}~\Msun$. The low average SFR indicate that most of the red objects are unlikely to be dusty star-forming galaxies. 
    
    \item  We stacked the \hii\ spectra of 823 galaxies hosting radio-bright AGNs at $z=0.74-1.45$, but did not obtain a detection of the average \hii\ emission signal. This yielded the $3\sigma$ upper limit of $24\times10^9~\Msun$ on the average \hi\ mass of the 823 galaxies, consistent with that of the main sample of blue galaxies without AGNs at $z\approx1$.
    
    \item Finally, we searched the spectra of all 13,596 DEEP2 galaxies (including red galaxies, objects containing AGNs, and low-stellar mass galaxies) of the sample for \hii\ emission at the expected redshifted \hii\ line frequency, at a spatial resolution of 90~kpc and velocity resolutions of $30-300$~\kmps. We find no evidence for individual redshifted \hii\ emission from any of the DEEP2 galaxies, obtaining $5\sigma$ upper limits of $\approx(3-25) \times 10^{11}~\Msun$ on the \hi\ masses of individual galaxies.
    \end{itemize}

\begin{acknowledgments}
We thank the staff of the GMRT who have made these observations possible. The GMRT is run by the National Centre for Radio Astrophysics of the Tata Institute of Fundamental Research. NK acknowledges support from the Department of Science and Technology via a Swarnajayanti Fellowship (DST/SJF/PSA-01/2012-13). All authors acknowledge the Department of Atomic Energy for funding support, under project 12-R\&D-TFR-5.02-0700. 
\end{acknowledgments}
\software{CASA \citep{McMullin07},   
calR \citep{calR},  AOFLAGGER \citep{Offringa12}, numpy \citep{harris2020array}, astropy \citep{astropy:2013,astropy:2018}, matplotlib \citep{Hunter:2007}, scipy \citep{2020SciPy-NMeth}, numba \citep{lam2015numba}, joblib \citep{joblib}}
\bibliography{bibliography.bib}

\bibliographystyle{aasjournal}	

\appendix

\section{The RMS noise on the individual \hii\ spectra}
\label{appndx:rmsnoise}

\begin{figure}[!b]
    \centering
 \includegraphics[width=0.6\linewidth]{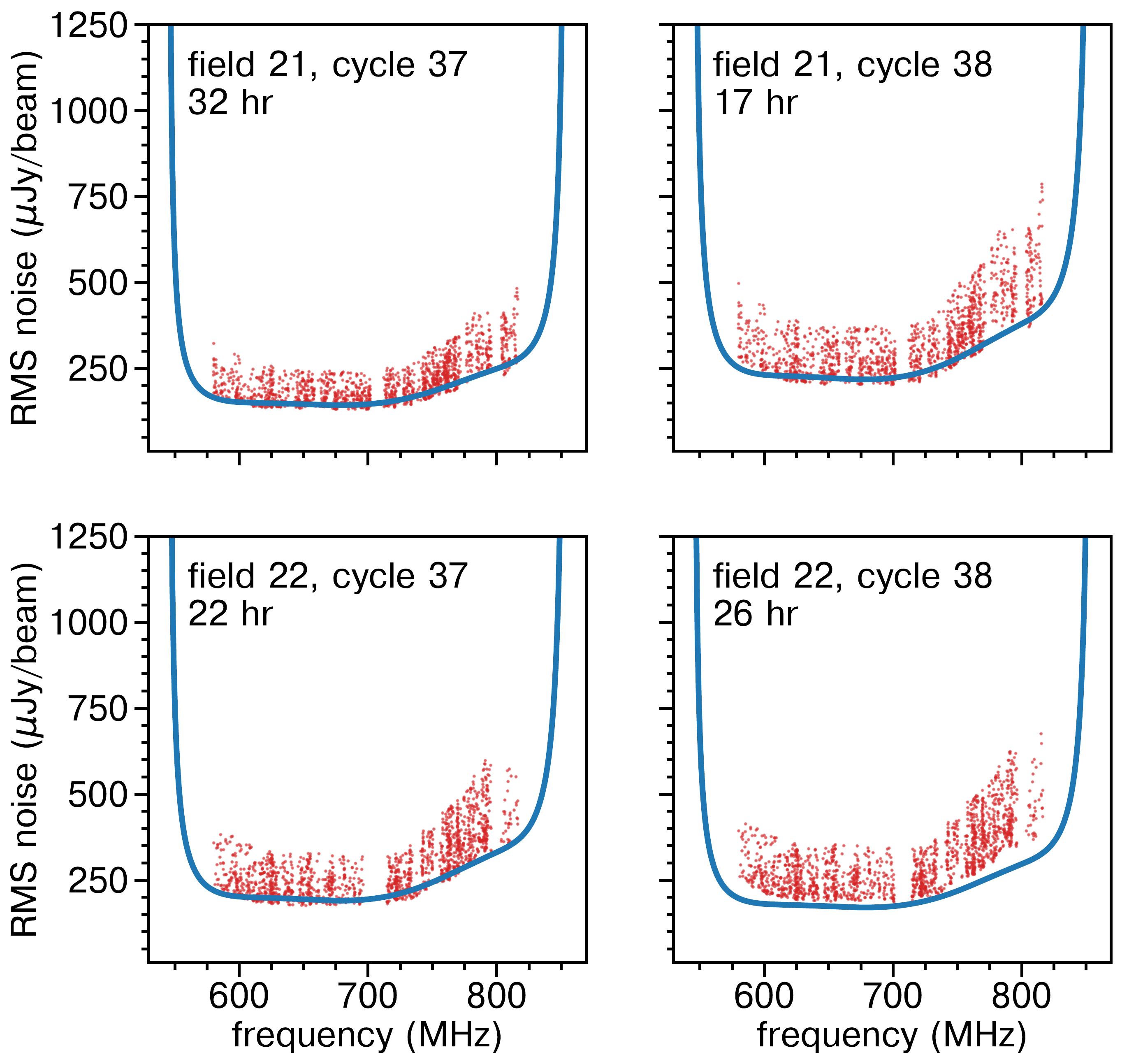}
 \caption{The spectral RMS noise on the \hii\ subcubes of galaxies in DEEP2 subfields 21 and 22. Observations in different GMRT cycles (35, 37, and 38) are shown in separate panels, with the red dots indicating the spectral RMS noise per 30~\kmps\ channel for each galaxy as a function of observing frequency. The blue curve in each panel shows the theoretical RMS noise for the amount of on-source time, taking into account the data that were excised, and also considering the spatial resolution of 90~kpc at which the \hii\ spectra were extracted. The panels list the on-source time for each DEEP2 subfield in each cycle. Note that the scatter by a factor of $\approx 2$ in the measured RMS noise values at any frequency is due to the location of the galaxies in the primary beam. The spectral RMS noise values obtained on the \hii\ spectra are consistent with the theoretical RMS noise, within the typical 10$\%$ uncertainty in the GMRT flux density scale.}
    \label{fig:rmsfield2}
\end{figure}
\begin{figure}
    \centering
 \includegraphics[width=0.9\linewidth]{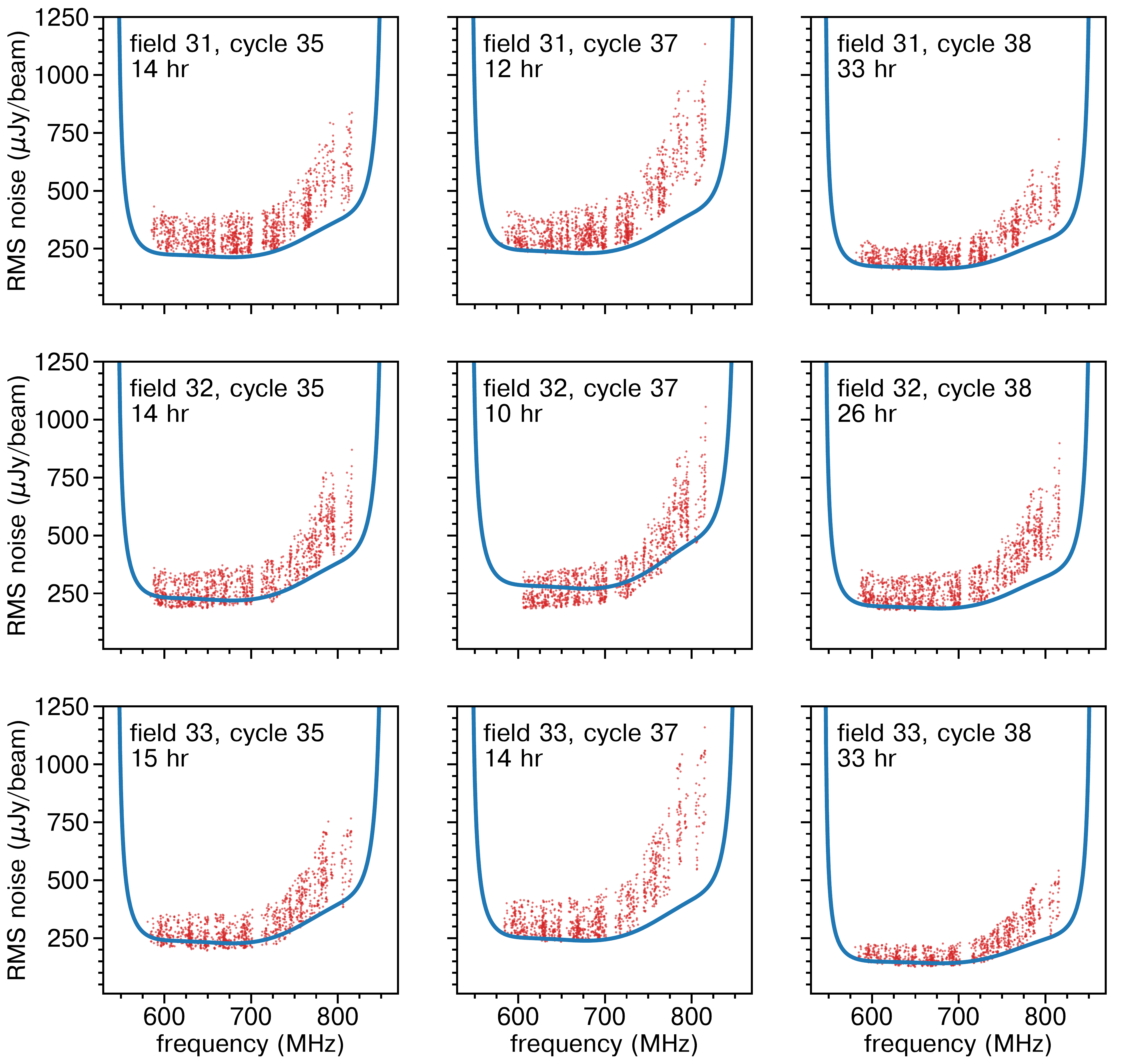}
 \caption{The spectral RMS noise on the \hii\ spectra of galaxies in DEEP2 subfields 31, 32, and 33. See the caption of  Fig.~\ref{fig:rmsfield2} for details.}
    \label{fig:rmsfield3}
\end{figure}
\begin{figure}
    \centering
 \includegraphics[width=0.9\linewidth]{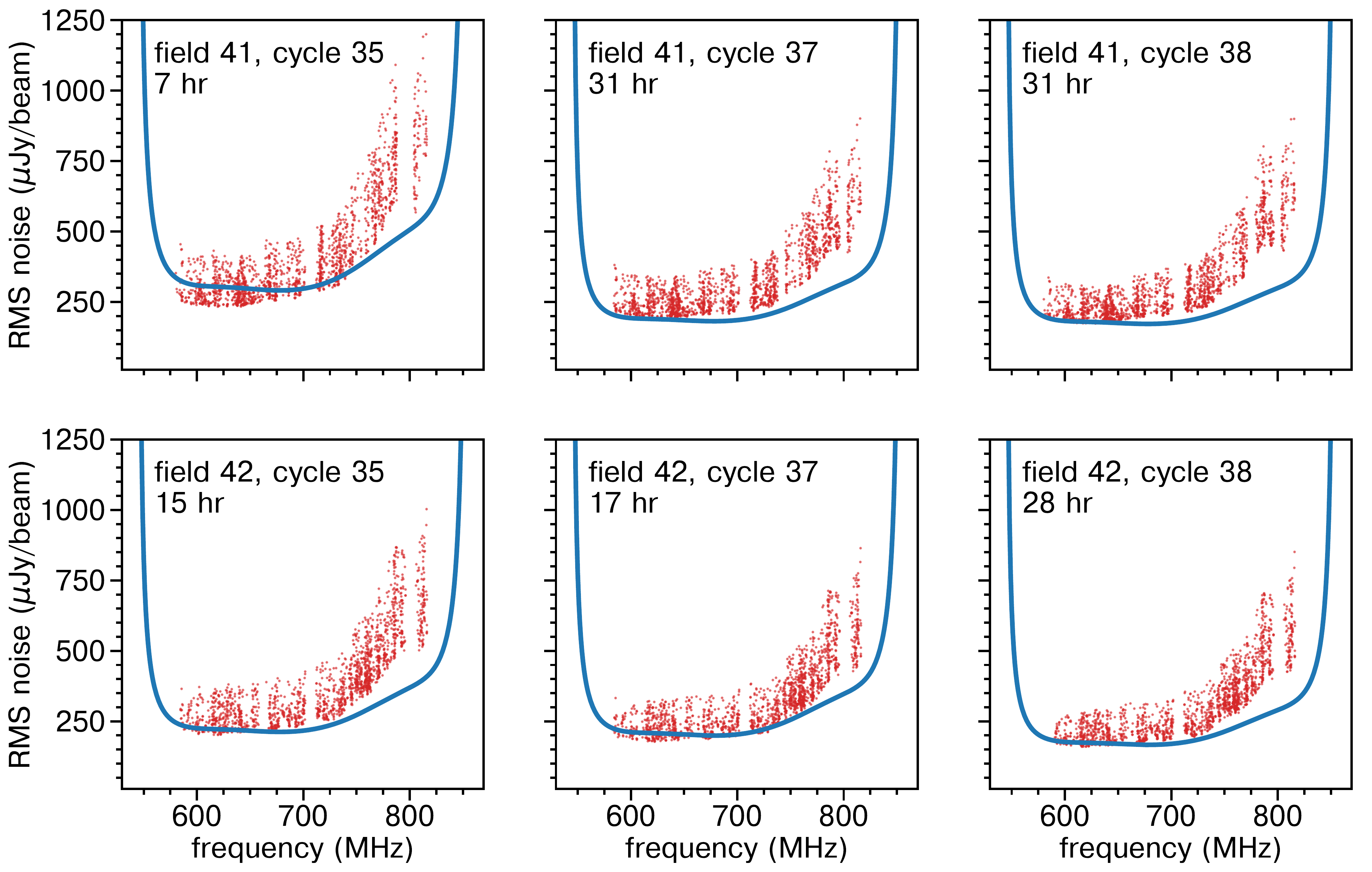}
 \caption{The spectral RMS noise on the \hii\ spectra of galaxies in DEEP2 subfields 41 and 42. See the caption of  Fig.~\ref{fig:rmsfield2} for details.}
    \label{fig:rmsfield4}
\end{figure}

\end{document}